\documentclass[10pt,journal,compsoc]{IEEEtran}

\usepackage{graphicx}
\usepackage{graphics}
\usepackage{subfigure}
\usepackage{multirow}
\usepackage{textcomp}
\usepackage{array}
\usepackage{xcolor}

\usepackage{algorithm}
\usepackage{algorithmic}
\usepackage[algo2e,linesnumbered,ruled]{algorithm2e}

\usepackage{amsthm}
\theoremstyle{definition}

\newtheorem{definition}{Definition}

\newtheorem{thm}{Theorem}

\SetKwProg{Fn}{Function}{}{end}

\SetAlgoSkip{}

\newcolumntype{M}[1]{>{\centering\arraybackslash}m{#1}}

\DeclareSymbolFont{symbolsC}{U}{txsyc}{m}{n}
\DeclareMathSymbol{\notniFromTxfonts}{\mathrel}{symbolsC}{61}

\ifCLASSOPTIONcompsoc
\usepackage[nocompress]{cite}
\else
\usepackage{cite}
\fi
\ifCLASSINFOpdf
\else
\fi
\usepackage{amsmath,amssymb,amsfonts}
\usepackage{bm}
\begin{document}
	%
	\title{Distributed Indexing Schemes for k-Dominant Skyline Analytics on Uncertain Edge-IoT Data}
	%
	%
	%
	%

	\author{Chuan-Chi~Lai,~\IEEEmembership{Member,~IEEE,}
		Hsuan-Yu~Lin,
		and~Chuan-Ming Liu,~\IEEEmembership{Member,~IEEE}
		\IEEEcompsocitemizethanks{
			\IEEEcompsocthanksitem{The preliminary result of this work was presented in WOCC 2021~\cite{9603141}.}
			\IEEEcompsocthanksitem{This research was supported by the Ministry of Science and Technology, Taiwan under the Grant Nos. MOST 109-2221-E-027-095-MY3 and MOST 110-2222-E-035-004-MY2. (Corresponding author: Chuan-Ming Liu.)}
			\IEEEcompsocthanksitem{Chuan-Chi Lai is affiliated to the Department of Information Engineering and Computer Science, Feng Chia University, Taichung 40724, Taiwan. (Email: chuanclai@fcu.edu.tw)}
			\IEEEcompsocthanksitem{Hsuan-Yu Lin is affiliated to the Tool Productivity Optimization Department, Taiwan Semiconductor Manufacturing Co., Ltd. (TSMC), Hsinchu 30078, Taiwan. (Email: danny30lin@gmail.com)}%
			\IEEEcompsocthanksitem{Chuan-Ming Liu is affiliated to the Department of Computer Science and Information Engineering, National Taipei University of Technology, Taipei 10618, Taiwan. (Email: cmliu@ntut.edu.tw)}
		}
	}
	
	%
	%

\markboth{Preprint submitted to IEEE Journal for publication}%
{Lai \MakeLowercase{\textit{et al.}}: Bare Demo of IEEEtran.cls for IEEE Communications Society Journals}
%



\IEEEtitleabstractindextext{%
	\begin{abstract}
		Skyline queries typically search a Pareto-optimal set from a given data set to solve the corresponding multiobjective optimization problem. As the number of criteria increases, the skyline presumes excessive data items, which yield a meaningless result. To address this curse of dimensionality, we proposed a $k$-dominant skyline in which the number of skyline members was reduced by relaxing the restriction on the number of dimensions, considering the uncertainty of data. Specifically, each data item was associated with a probability of appearance, which represented the probability of becoming a member of the $k$-dominant skyline. As data items appear continuously in data streams, the corresponding $k$-dominant skyline may vary with time. Therefore, an effective and rapid mechanism of updating the $k$-dominant skyline becomes crucial. Herein, we proposed two time-efficient schemes, Middle Indexing (MI) and All Indexing (AI), for $k$-dominant skyline in distributed edge-computing environments, where irrelevant data items can be effectively excluded from the compute to reduce the processing duration. Furthermore, the proposed schemes were validated with extensive experimental simulations. The experimental results demonstrated that the proposed MI and AI schemes reduced the computation time by approximately 13\% and 56\%, respectively, compared with the existing method.
	\end{abstract}
	
	\begin{IEEEkeywords}
		Pareto-optimal, $k$-dominant skyline, skyline, uncertain data, data streams	
\end{IEEEkeywords}}

\maketitle

\IEEEdisplaynontitleabstractindextext

%
\IEEEpeerreviewmaketitle

\IEEEraisesectionheading{\section{Introduction}\label{sec:introduction}}
\IEEEPARstart{S}{kyline} is an efficient analysis tool~\cite{borzsony2001skyline,10.5555/1287369.1287394,1260846} for solving multiobjective optimization and multi-criteria decision-making problems in the big data of the Internet of Things (IoT). 
It also has been widely studied and applied in numerous applications, such as location-based services~\cite{AL_Jawarneh2020,9511122}, transportation~\cite{CHAOLONG2016719,9101829}, mobile computing~\cite{9091902,9354847}, Internet of Mobile Things~\cite{8681541,9380540,9454452}, and social networks~\cite{9101805,9509754}. 
As no singularly best answer exists for multi-criteria decision-making applications, the skyline (or Pareto-optimal front) has become a popular approach. Skyline query can assist users to determine the results that fulfills their multi-criteria needs, and thus numerous efficient skyline query methods~\cite{tan2001efficient,10.1145/1327452.1327492,zhang2015efficient,huang2018efficient} have been developed. 

\begin{figure}[!t]
	\centering
	\includegraphics[width=.65\columnwidth]{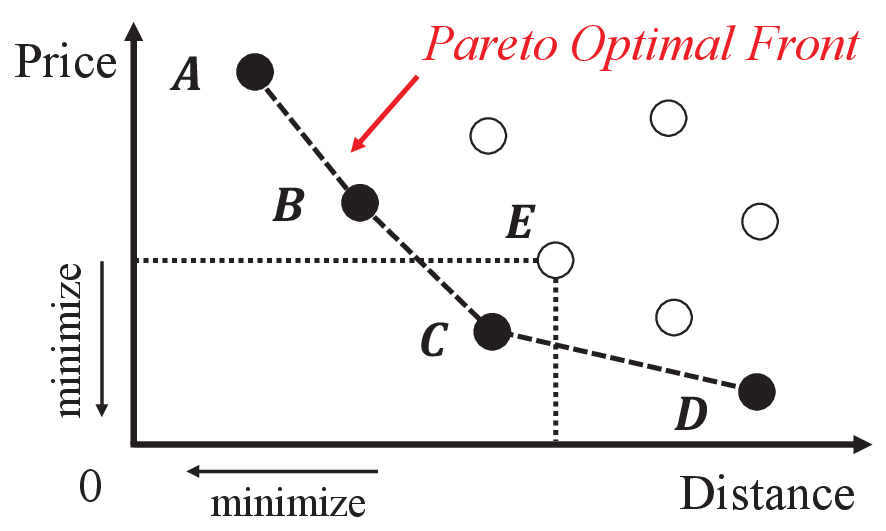}
	\caption{Example of two-dimensional skyline for selecting a hotel}
	\label{fig:fig1}
\end{figure}

	Several well-known location-based services such as Agoda, Hotel.com$^{\text{TM}}$, and trivago$^{\text{TM}}$, have widely applied skyline queries. To reserve a hotel near a given venue, attendees generally consider the following two factors: distance and price.
The attendee may obtain a set of candidate hotels, which is the skyline (or Pareto-optimal front) recommended by these location-based services. As depicted in Fig.~\ref{fig:fig1}, the attendee can decide by selecting one of the four hotels from a recommended skyline set of $\{A,B,C,D\}$.

	However, as the volume and dimensionality of data increase, the size of the recommended skyline set tends to explode~\cite{10.1145/322092.322095,chan2006finding}, known as the curse of dimensionality. 
The expected size of the skyline of $n$ random vectors/data, with a randomly selected criterion, is $O((\log n)^{d-1})$ in dimensions $d$~\cite{10.1145/322092.322095}.
The exponential dependence of skyline on $d$ renders it theoretically useless for IoT analytic applications, except in extremely low dimensions.
The quotation information of computer components are listed in Table~\ref{table:quotation_information}. shows the quotation information of computer components. The values in this table are normalized to $[1,9]$ space, wherein a lower value represents a better ranking of the corresponding computer components. Suppose a customer intends to purchase a computer, but owing to budget constraints, the best equipment cannot be readily determined. In this case, we can use a skyline query to search options that match the customer's criteria. However, if $d=7$ in this example, the skyline result will be $\{u_1,u_2,u_3,u_4,u_5\}$, which is not meaningful for the customer.

\begin{table*}[!t]
	\centering
	\caption{Quotation information of computer components}
	\label{table:quotation_information} 
	\begin{tabular}{|l|lllllll|}
		\hline
		\textbf{Computer} & \textbf{CPU} & \textbf{Memory} & \textbf{HDD} & \textbf{VGA} & \textbf{Motherboard} & \textbf{Power Supply} & \textbf{Blu-ray}\\ \hline 
		$u_1$ & 4 & 8 & 2 & 2 & 8 & 6 & 4\\ 
		$u_2$ & 9 & 8 & 7 & 2 & 1 & 5 & 8\\ 
		$u_3$ & 3 & 2 & 4 & 5 & 6 & 1 & 7\\ 
		$u_4$ & 6 & 3 & 4 & 8 & 1 & 2 & 3\\ 
		$u_5$ & 5 & 7 & 9 & 5 & 7 & 3 & 1\\ 
		\hline 
	\end{tabular}
\end{table*}

To overcome this curse of dimensionality, we propose a variant of skyline--the $k$-dominant skyline. By relaxing the $d$-dominance to $k$-dominance, the $k$-dominant skyline can obtain a smaller and more manageable set of maxima (or minima), because if $k<d$, more data items can be eliminated from the skyline. 
Several methods~\cite{chan2006finding,siddique2009k,siddique2012efficient,tian2014efficient,lee2017efficient} have discussed $k$-dominant skyline processing based on the certain/deterministic data model, but not all data are certain/deterministic in practical IoT applications. Owing to the aging of the sensor, the data transmitted by the sensor is uncertain. For instance, the location information from the global positioning system (GPS) sensor often contains errors.

	Unlike traditional methods for processing the $k$-dominant skyline on deterministic data, the method of calculating the $k$-dominant skyline on uncertain data~\cite{1617375,pei2007probabilistic,Prabhakar2009,Wang2013} is much more complicated~\cite{li2019parallel}. 
Although certain recent studies~\cite{8731646,9162084} considered spatial query processing with edge computing, the $k$-dominant skyline processing method based on edge computing environment with uncertain data has been seldom discussed.

This research gap inspired us to propose a method for rapidly evaluating and updating the probability of a $k$-dominant skyline in an uncertain edge-IoT data stream environment.
In such an environment, each data item bears a specific probability of becoming a $k$-dominant skyline. Owing to the continuous inflow and outflow of data in the system, the $k$-dominant probability of each item varies with time time, i.e., the $k$-dominance probability of each data item must be updated at each instant. 
Therefore, calculating and updating the probability of the $k$-dominant skyline in an environment of uncertain data flow requires an enormous amount of computation.

In this study, we propose two distributed indexing schemes and apply them to the edge-computing environment to improve the performance of $k$-dominant skyline query processing. Through the proposed distributed indexing scheme, the system can effectively filter out irrelevant information, reduce unnecessary calculations, and thus, accelerate the computation speed. The contributions of the present study are described as follows.
\begin{itemize}
	\item This study explored $k$-dominant skyline query processing over uncertain IoT data streams in edge-computing environments.
	\item We proposed and deployed two new distributed indexing schemes, \emph{Middle Indexing} (MI) and \emph{All Indexing} (AI), on multiple edge computing nodes to prune irrelevant information in a distributed manner, and thus, effectively improve the performance of the $k$-dominant skyline derivation on uncertain IoT data streams.
	\item The simulation results indicated that the proposed MI and AI significantly improved the performance of computing $k$-dominant skyline in terms of computation time by approximately 13\% and 56\%, respectively.
	\item Compared with the existing method ~\cite{li2019parallel}, the proposed AI scheme is more suitable for processing high-dimensional uncertain data.
\end{itemize}

The remainder of the paper is organized as follows. The current  state-of-the-art skyline models are reviewed in Section~\ref{sec:related_work}. The preliminaries and the problem statements of the current research are expressed in Section~\ref{sec:problem}. Thereafter, the proposed approaches with algorithms and examples are presented in 
Section~\ref{sec:proposed_approach}. The simulation results are discussed in in Section~\ref{sec:simulation}. Finally, the conclusions of this study are summarized in Section~\ref{sec:conclusion}.

\section{Related Work}
\label{sec:related_work}
The tools suitable for solving multiobjective optimization problems vary across domains. For example, to solve the considered multiobjective optimization problems in the fields of radio resource management and wireless communications, game theoretic techniques~\cite{9846949,10016754,9040594} and genetic algorithm-based methods~\cite{10036140,10039683} are widely applied. However, the majority of literature in wireless networks focus on realizing optimal performance, and therefore, generate only one Pareto candidate. Conversely, in the fields of database management, location-based services, and recommendation system, $k$-nearest neighbor~\cite{10.1007/978-3-540-28651-6_80}, skyline~\cite{9186333,10064018}, and top-$k$~\cite{8731646,9027928} are more popular because most applications in these domains aim at recommending sets of Pareto candidates to the decision-makers. Based on these recommendations of the Pareto candidate sets, the decision-maker derives the final decision. As the present study is closely related to location-based services and recommendation systems, we did not review the literature on wireless communication and networking issues.

Based on the perspective of data engineering,~\cite{borzsony2001skyline} proposed the idea of skyline queries to extract multicriteria decisions.
Tan \emph{et al.}~\cite{tan2001efficient} proposed two indexing schemes, Bitmap and B$^+$-tree, for improving the performance of skyline processing. Thereafter, certain studies~\cite{zhang2015efficient}~\cite{huang2018efficient} proposed parallel processing solutions for skyline processing. Zhang \emph{et al.}~\cite{zhang2015efficient} used the MapReduce framework for parallel processing of the skyline, and proposed three algorithms: \emph{aggressive partition-aware filtering} (APF), \emph{professional partition-aware filtering} (PPF) and \emph{partial-presort grid-based partition skyline} (PGPS). APF and PPF could filter numerous items in a distributed manner, and PGPS utilized the shuffle processing stage of MapReduce to effectively calculate the skyline from a large amount of data. Huang \emph{et al.}~\cite{huang2018efficient} proposed a \emph{cost-based algorithm} (CA) according to skyline views to improve the efficiency of the skyline query. The CA algorithm used MapReduce to accelerate the generation of the skyline view (view set).

\begin{table*}[!t]
	\centering
	\caption{Comparisons Of Existing Works}
	\label{table:comparison_of_methods} 
	\begin{tabular}{|l|l|l|l|}
		\hline
		\textbf{Methods} & \textbf{Query Type} & \textbf{Data Type} & \textbf{Edge Computing}\\ \hline 
		\cite{tan2001efficient}~\cite{zhang2015efficient}~\cite{huang2018efficient} & Skyline & Deterministic & \texttimes \\ 
		\cite{ding2011efficient}~\cite{liu2015effective}~\cite{park2015processing} & Skyline & Uncertain & \texttimes \\ 
		\cite{chan2006finding}~\cite{siddique2009k}~\cite{siddique2012efficient}~\cite{tian2014efficient}~\cite{lee2017efficient} & $k$-Dominant Skyline & Deterministic & \texttimes \\ 
		\cite{li2019parallel} & $k$-Dominant Skyline & Uncertain & \texttimes \\ 
		\textbf{Present study} & \textbf{$\bm{k}$-Dominant Skyline} & \textbf{Uncertain} & \textbf{\checkmark} \\ 
		\hline 
	\end{tabular}
\end{table*}

To support uncetain data, research related to the skyline has been discussed~\cite{ding2011efficient,liu2015effective,park2015processing}. Ding and Jin~\cite{ding2011efficient} evaluated the skyline of uncertain data on a distributed architecture and reduced the computational burden of each distributed node using \emph{probabilistic R-tree} (PR-tree). Liu and Tang~\cite{liu2015effective} proposed an \emph{effective probabilistic skyline update} (EPSU) method with an augment R-tree structure: \emph{SW-tree}. With EPSU, uncertain data from the input data stream can be effectively managed, and the time and space required to calculate the probabilistic skyline can be reduced. Park~\emph{et al.}~\cite{park2015processing} used MapReduce and proposed the \emph{probabilistic skyline algorithm by quadtree partitioning with MapReduce} (PS-QP-MR) algorithm based on quadtree to process discrete and continuous uncertain data. The authors added a filtering stage to the PS-QP-MR, denoted as the PS-QPF-MR algorithm, to effectively distribute the instances of data items for improving the efficiency of probabilistic skyline queries.

As the number of criteria (or dimensions) increases, it is less likely that one data item dominates another data item, which causes the inclusion of excessive members of the skyline set, and not all criteria of the skyline Set are required by the user. To this end, \cite{chan2006finding} proposed the concept of $k$-dominant skyline, which reduced the number of skyline members by relaxing the number of criteria considered from $d$ to $k$ in $d$-dimensional data, where $k<d$. Siddique and Morimoto~\cite{siddique2009k} proposed the \emph{sort-filtering} method, which is suitable for large-scale high-dimensional data and can reduce the time required to calculate the $k$-dominant skyline query. In particular, they proposed the \emph{domination power} method~\cite{siddique2012efficient}, which effectively reduced the number of comparisons required to calculate the $k$-dominant skyline by sorting each item in the sliding window in a specific manner. Under the framework of MapReduce, Tian~\emph{et al.}~\cite{tian2014efficient} proposed to the use of a \emph{point-based bound tree} (PB-tree) to partition the data space and perform parallel calculations. Using the PB-tree method, the workload of the $k$-dominant skyline can be simply distributed and efficiently calculated. As such, four lemmas are proposed to reduce the time required to select candidates and trim false positives in the calculation process of $k$-dominant skyline~\cite{lee2017efficient}. However, the above methods did not considered uncertain data.

Under the environment of uncertain data streams, Li~\emph{et al.}~\cite{li2019parallel} proposed the \emph{parallel $k$-dominant skyline with capability index} (PKDS-CI) approach to rapidly compare the $k$-dominant relationship between two data items. 
This method initially normalizes each dimension of data and sorts the data according to their normalized values in all dimensions in ascending order.
Thereafter, the product of the first $k$ small values were sorted, the product of the first $k$ large values of each incoming item was determined as the key value, and the sorted results were stored in a table. Based on this normalized and sorted index table, the author proposed the CI theorem to establish that the index effectively filtered out unnecessary calculations, thereby accelerating the update of the $k$-dominant skyline. 

In this study, we propose two schemes, MI and AI, based on a distributed edge environment, to process $k$-dominant skyline queries over uncertain edge-IoT data streams. To emphasize the novelty of this work, the proposed schemes and the aforementioned methods are comparatively summarized in Table~\ref{table:comparison_of_methods}. Furthermore, the proposed MI and AI schemes have been compared with the PKDS-CI in the simulation section.

\section{Problem Description}
\label{sec:problem}
In this section, we sequentially introduce the preliminary assumptions, system architecture, and problem statement. The notations used in this research are listed in Table~\ref{table:notations}.

\begin{table}[!t]	
	\centering
	\caption{Notations Used Throughout This Paper}
	\label{table:notations}  
	\begin{tabular}{c|p{5.7cm}}
		\hline
		\textbf{Symbol} & \textbf{Description} \\ \hline
		$d$				&  Total number of data dimensions  \\ 
		$k$				&  The target number of dominant dimensions \\
		$U$				&  A set of uncertain data items \\ 
		$u$				&  An uncertain data item  \\ 
		$S$				&  The attribute space of $d$-dimensional uncertain data \\ 
		$u\cdot s_j$	&  The $j$-th attribute value of $u$ \\ 
		$m$	            &  The number of edge computing nodes \\ 
		$N_H$	        &  The cloud/header server \\ 
		$N_e$	        &  The $e$-th edge computing node, where $e=1,2,\dots,m$ \\ 
		$SW$			&  Sliding window  \\ 
		$SW(t)$			&  The instance of $SW$ at time $t$ \\ 
		$|SW(t)|$		&  The size of $SW$ at time $t$  \\ 
		$|SW|_{\max}$	&  The predefined maximum size of $SW$  \\ 
		$\mathbb{P}(u)$	& The occurrence probability of $u$ \\ 
		$\mathbb{P}_{k-{\rm sky}}(u)$	&  The probability of $u$ being the $k$-dominance skyline\\ 
		$u_{\rm new}$	&  A new data item coming into $SW$ \\ 
		$u_{\rm old}$	&  An old data item leaving $SW$ \\ 
		$u_{\rm sw}$	&  A remaining data item in $SW$ \\ 
		$SORTED(U)$		&  The normalized and sorted dataset of $U$  \\ 
		$SORTED(u)$		&  The normalized and sorted tuple of $u$ \\ 
		$SORTED(u)[\alpha]$	&  The value stored in the $m$-th position of $SORTED(u)$, where $\alpha=0,1,\dots,d-1$ \\ 
		$u_{\min}(k)$	&  The first selected index position \\
		$u_{\max}(k)$	&  The second selected index position \\
		$MI_{\min}(u,k)$	&  The value of $SORTED(u)\left[u_{\min}(k)\right]$\\ 
		$MI_{\max}(u,k)$	&  The value of $SORTED(u)\left[u_{\max}(k)\right]$\\ 
		$MIT_{\min}(k)$	&  A sorted index table recording the remaining data items in $SW$ in ascending order of $MI_{\min}(u,k)$ \\ 
		$MIT_{\max}(k)$	&  A sorted index table recording the remaining data items in $SW$ in descending order of $MI_{\max}(u,k)$ \\ 
		$AIT_{\min}$	&  The remaining set of data items after updating $\mathbb{P}_{k-{\rm sky}}(u_{\rm sw})$ by using the proposed all-indexing scheme \\ 
		$AIT_{\max}$	&  The remaining set of data items after updating $\mathbb{P}_{k-{\rm sky}}(u_{\rm new})$ by using the proposed all-indexing scheme \\ \hline
	\end{tabular}
\end{table}

\subsection{Preliminary Assumptions}

In this subsection, we define the essential considerations and parameters of this work. Data with uncertainty are called uncertain data~\cite{1617375}~\cite{pei2007probabilistic}, which exists in several applications. For instance, multiple temperature sensors may be installed at the same location, and the temperature measured by each sensor may vary, thereby causing uncertainty in the data~\cite{Prabhakar2009}. This type of data uncertainty is common in environments such as environmental testing and location services. The mathematical definition of uncertain data is expressed as follows:

%

\begin{definition}[\textbf{Uncertain Data}]
	\label{def:DPDM}
	Given a $d$-dimensional space $S=\{s_1,s_2,\dots,s_d\}$, a set of uncertain data $U=\{u_1,u_2,\dots,u_n\}\in S$, and $u_i \cdot s_j$ representing the $j$-th dimensional value of $u_i$, where $i=1,2,\dots,n$ and $j=1,2,\dots,d$, the occurrence probability of uncertain data item $u_i$ can be denoted as $\mathbb{P}(u_i)$.
\end{definition} 

An example of an uncertain dataset containing five data items is presented in Table~\ref{table:uncertain_data_set}, wherein each data item contains four attributes and a probability value. In this example, the attribute values of the data item $u_1$ in four dimensions are 10,3,4, and 6, respectively; the occurrence probability of $u_1$ is $\mathbb{P}(u_1)=0.2$.

Owing to the continuous flow of data stream into the system, copious amounts of data are accumulated.
Generally, each datapoint is time-stamped and becomes outdated after a period of time. As these outdated data may provide unimportant information, they must be filtered out to ensure that they do not affect the correctness of the calculation. Because of the infiniteness of the data stream, all data cannot be calculated. As such, the sliding window model can identify the data of interest by filtering out the data that may affect the accuracy of the calculation, which considerably reduces the computation time. Thus, this study adopted the count-based sliding window, defined as follows:
\begin{definition}[\textbf{Count-Based Sliding Window}]
	\label{def:cb_sw}
	A sliding window at time $t$ is denoted as $SW(t)$. The maximum size of a sliding window is denoted as $|SW|_{\max}$. The size of the sliding window at time $t$ is denoted as $|SW(t)|$. At any instant, $|SW(t)|$ does not exceed the maximum size $n$, i.e., $|SW(t)|\leq |SW|_{\max}, \forall t$. In addition, the sliding window handles the data items in a first-in-first-out manner.
\end{definition} 


Herein, we assumed that $|SW|_{\max}=3$ and one new data item flows into the system at each instant, e.g., $u_1$ at $t=1$, $u_2$ at $t=2$, and so forth. In such a scenario, the sliding window will be satiated if $t\geq 3$ and the oldest data item is removed before inserting the new data item. A corresponding example is presented in Table~\ref{table:sliding_window} to exhibit the variations in the sliding window from $t=1$ to $t=5$.

To search the $k$-dominant skyline, the system should evaluate the dominant relationship between multiple uncertain items. The dominate is defined as follows:
\begin{definition}[\textbf{Dominate}]
	\label{def:dominate}
	Given two different data items, $u_a ,u_b\in U$. Item $u_a$ dominates item $u_b$, denoted as $u_a\prec u_b$, iff $u_a \cdot s_j \leq u_b \cdot s_j, \forall j=1,2,\dots,d$, and $u_a \cdot s_{j'}<u_b \cdot s_{j'}, \exists j'\in \{1,2,\dots,d\}$. 	
\end{definition}

\begin{table}[!t]
	\centering
	\caption{Example Of An Uncertain Data Set}
	\label{table:uncertain_data_set} 	
	\begin{tabular}{|l|l|l|l|l|l|}
		\hline
		\textbf{Item} & \textbf{Attr1}& \textbf{Attr2} & \textbf{Attr3} & \textbf{Attr4} & \textbf{Probability}\\
		\hline
		$u_1$ & 10 & 3 & 4 & 6 & 0.2 \\ 
		$u_2$ & 9 & 8 & 5 & 9 & 0.4 \\ 
		$u_3$ & 2 & 10 & 4 & 4 & 0.5 \\ 
		$u_4$ & 5 & 2 & 3 & 8 & 0.1 \\ 
		$u_5$ & 7 & 6 & 4 & 6  & 0.8 \\
		\hline
	\end{tabular}
\end{table}

\begin{table}[!t]
	\centering
	\caption{Example Of A Sliding Window}
	\label{table:sliding_window} 
	\begin{tabular}{|l|l|l|}
		\hline
		\textbf{Time Slot} & \textbf{Sliding Window} & \textbf{Size} \\ \hline
		1	&  $SW_1=\{u_1\}$  &  $|SW_1|=1$    \\ 
		2	&  $SW_2=\{u_1,u_2\}$  &  $|SW_2|=2$    \\ 
		3	&  $SW_3=\{u_1,u_2,u_3\}$  &  $|SW_3|=3$    \\ 
		4	&  $SW_4=\{u_2,u_3,u_4\}$  &  $|SW_4|=3$    \\ 
		5	&  $SW_5=\{u_3,u_4,u_5\}$  &  $|SW_5|=3$    \\ \hline
	\end{tabular}
\end{table}

Considering the example in Table~\ref{table:uncertain_data_set}, if $u_4$ is not worse than $u_2$ in all attributes (or dimensions) and $u_4$ outperforms $u_2$ in at least one attribute, so we can say $u_4$ dominates $u_2$, denoted as $u_4\prec u_2$.

In general, the dominant relations between different data items have a transitivity property. This phenomenon is called the \emph{transitivity of domination} and is defined as follows:
\begin{definition}[\textbf{Transitivity of Domination}]
	\label{def:transitivity}
	Given three different data items, $u_a, u_b, u_c\in U$. If $u_a\prec u_b$ and $u_b\prec u_c$, such that $u_a\prec u_c$. 
\end{definition} 


Based on Definitions~\ref{def:dominate} and~\ref{def:transitivity}, $k$-dominant can be defined as
\begin{definition}[\textbf{$\bm{k}$-Dominate}]
	\label{def:k_dominance}
	Given two different data items, $u_a, u_b\in U$, $u_a$ $k$-dominates $u_b$, denoted as $u_a\prec_k u_b$, iff the following two conditions hold simultaneously:
	\begin{enumerate}
		\item $u_a\cdot s_j\leq u_b\cdot s_j, \forall s_j\in S'$, where $\exists S'\subseteq S, |S'|\geq k$;
		\item $u_a\cdot s_{j'}<u_b\cdot s_{j'}, \exists s_{j'}\in S$.
	\end{enumerate}
\end{definition} 

Referring to the exemplary case in Table~\ref{table:uncertain_data_set}, the data item $u_1$ was not worse than $u_2$ in $\mathsf{Attr2}$, $\mathsf{Attr3}$ and $\mathsf{Attr4}$ attributes and outperforms $u_2$ in at least one attribute such as attribute $\mathsf{Attr2}$, such that $u_1$ 3-dominates $u_2$, denoted as $u_1\prec_3 u_2$.

According to Definitions~\ref{def:dominate} and~\ref{def:k_dominance}, $k$-dominance is a relaxing variant of dominance (also called $d$-dominance) and $k<d$. However, such a relaxation influences $k$-dominance to violate the transitivity of domination. Considering the example in Table~\ref{table:uncertain_data_set}, if $k=2$, $u_1\prec_2 u_3$ and $u_3\prec_2 u_1$, this phenomenon is called the \emph{cyclic dominance} (CD) relationship. As most existing skyline query methods follow the transitivity of dominance, they cannot be directly applied to the $k$-dominant skyline query. 

Therefore, based on these assumptions and definitions, the $k$-dominant skyline can be defined as follows:
\begin{definition}[\textbf{$\bm{k}$-Dominant Skyline}]
	\label{def:k_dominant_skyline}
	Given a $d$ dimensional space $S$, for data item $u \in S$, none of data items $u'\in S$ can $k$-dominates $u$, such that $u$ is the $k$-dominant skyline, expressed as
	\begin{equation*}
		U_{k\text{-sky}} = \{u|\nexists u'\prec_k u, u'\neq u, u\in S, u'\in S\}.
	\end{equation*}
\end{definition} 

In Table~\ref{table:uncertain_data_set}, none of data items can $k$-dominate $u_3$ and $u_4$, such that $\{u_3,u_4\}$ represents the $k$-dominant skyline.

In an environment with uncertain data, each data item manifests a probability of composing the $k$-dominant skyline. Therefore, the probability of data item $u$ representing the $k$-dominant skyline is defined as 
\begin{definition}[\textbf{Probability of Being $\bm{k}$-Dominant Skyline}]
	\label{def:probability_of_being_k_dominant_skyline}
	According to the aforementioned assumptions and definitions, the probability of data item $u$ being the $k$-dominant skyline can be expressed as
	\begin{equation}\label{eq:obj_dominant_probability}
		\mathbb{P}_{k\text{-sky}}(u)=\mathbb{P}(u)\times\prod_{u'\in S,u'\prec_k u}\left(1-\mathbb{P}(u')\right),
	\end{equation}
	where $u, u'\in SW$.
\end{definition}

\begin{figure}[!t]
	\centering
	\includegraphics[width=\columnwidth]{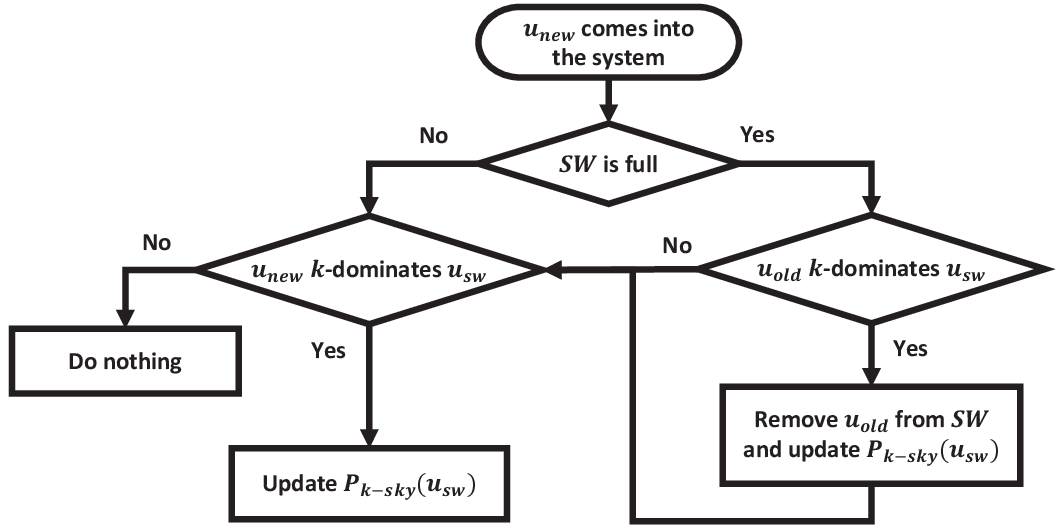}
	\caption{Procedure flowshart of updating $u_{\text{sw}}$, where $u_{\text{sw}}=\{u|\forall u\in SW\setminus\{u_{\text{new}}\}\}$}
	\label{fig:fig2}
\end{figure}

If a new data item $u$ flows into the system, denoted as $u_{\text{new}}$, the system will derive the probability of $u_{\text{new}}$ as the $k$-dominant skyline, $\mathbb{P}_{k\text{-sky}}(u_{\text{new}})$, using~\eqref{eq:obj_dominant_probability}. Moreover, the probability of each data item remaining in the sliding window must be updated. If a data item $u$ becomes outdated, denoted as $u_{\text{old}}$, the system will ignore $u_{\text{old}}$ and eliminate it from $SW$. If certain data items are still valid after adding $u_{\text{new}}$ into $SW$, these data items are denoted as a set $U_{\text{sw}}=\{u|\forall u\in SW\setminus\{u_{\text{new}}\}\}$. For simplicity, a data item $u\in U_{\text{sw}}$ is denoted as $u_{\text{sw}}$. The procedure of updating $\mathbb{P}_{k\text{-sky}}(u_{\text{sw}})$ is illustrated in Fig.~\ref{fig:fig2}.
The process of updating the probability of a data item $u_{\text{sw}}$ as the $k$-dominant skyline follows the definition stated below:
\begin{definition}[\textbf{$\bm{k}$-Dominant Skyline Probability Update}]
	\label{def:probability_of_being_k_dominant_skyline_update}
	When a new data item $u_{\text{new}}$ flows into the system, if the sliding window $SW$ is already full, the system should remove an old data item $u_{\text{old}}$ from $SW$ in advance. After eliminating $u_{\text{old}}$ from $SW$, the system updated each data item remaining in $U_{\text{sw}}$, denoted as $u_{\text{sw}}$, according to	
	\begin{align}\label{eq:obj_dominant_probability_update1}
		\mathbb{P}_{k\text{-sky}}(u_{\text{sw}})=&
		\mathbb{P}_{k\text{-sky}}(u_{\text{sw}})/(1-\mathbb{P}\left(u_{\text{old}})\right),\nonumber\\ 
		& \quad\text{ if } u_{\text{old}}\prec_k u_{\text{sw}}\wedge |SW|=|SW|_{\max},
	\end{align}
	where $U_{\text{sw}}=\{u|\forall u\in SW\setminus\{u_{\text{new}}\}\}$.
	If $SW$ contains free space/slot or the aforementioned procedure of elimination and update have been executed, the system adds $u_{\text{new}}$ into $SW$ and derives $\mathbb{P}_{k\text{-sky}}(u_{\text{new}})$ using~\eqref{eq:obj_dominant_probability}. Thereafter, the system updates each data item $u_{\text{sw}}$ (each data item $u$ remaining in $U_{\text{sw}}$) using	
	\begin{align}\label{eq:obj_dominant_probability_update2}
		\mathbb{P}_{k\text{-sky}}(u_{\text{sw}})=&
		\mathbb{P}_{k\text{-sky}}(u_{\text{sw}})\times \left(1-\mathbb{P}(u_{\text{new}})\right),\nonumber\\ 
		& \quad\text{ if } u_{\text{new}}\prec_k u_{\text{sw}}\wedge |SW|<|SW|_{\max}.
	\end{align}
\end{definition}

\subsection{System Architecture}
\begin{figure}[!t]
	\centering
	\includegraphics[width=\columnwidth]{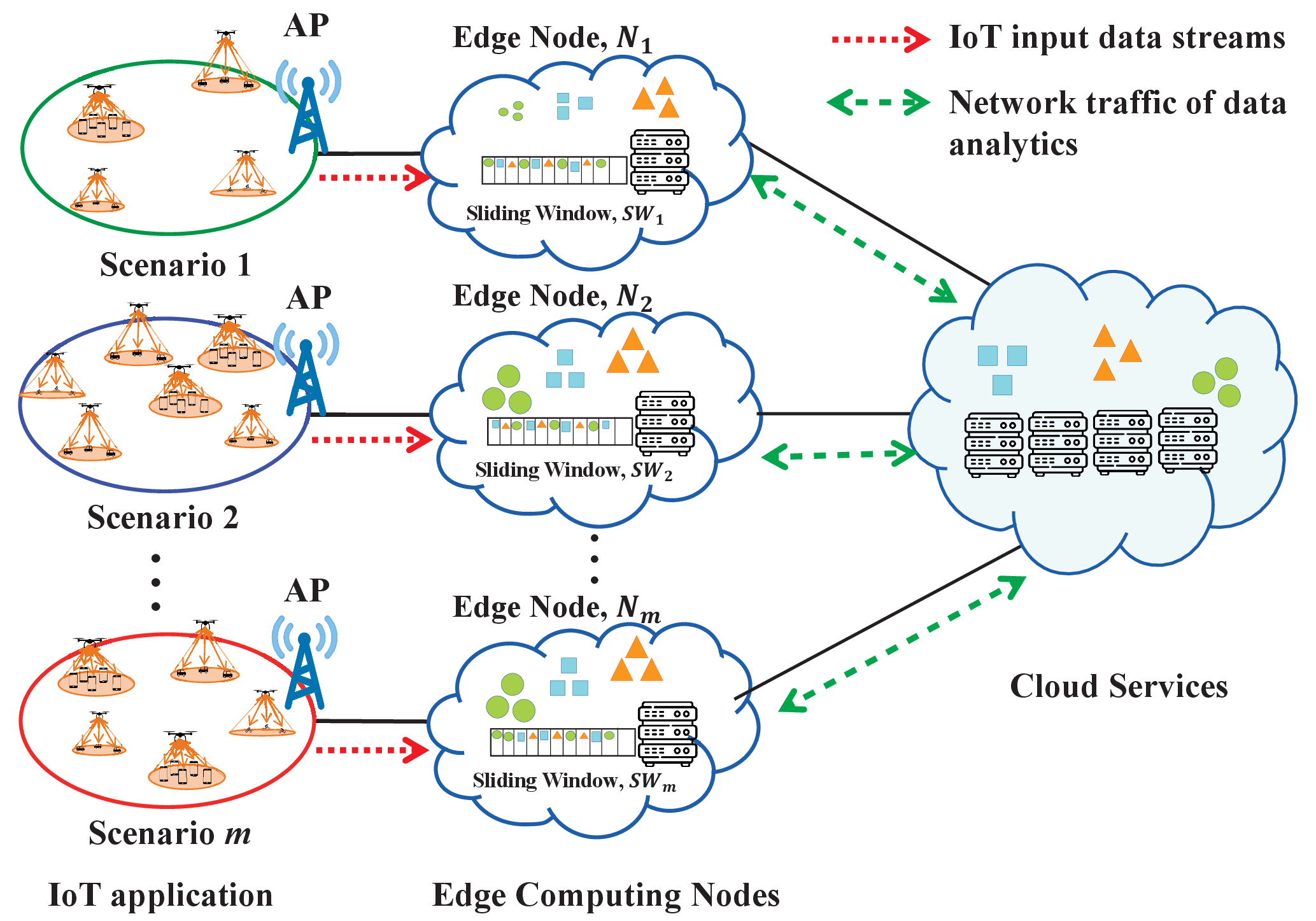}
	\caption{Distributed edge-IoT system architecture of considered herein}
	\label{fig:system_model}
\end{figure}

Herein, we considered a distributed edge computing environment with uncertain IoT sensing data sources, as displayed in Fig.~\ref{fig:system_model}. We utilized this architecture to devise a parallel and distributed computing framework for efficiently processing $k$-dominant skyline queries over multiple uncertain IoT data streams. There are $m$ edge computing nodes, $N_1, N_2, \dots, N_m$, with adequate computing resources and one main cloud server, $N_H$. All the data coming into edge nodes are uncertain IoT sensing data streams. Each edge node, $N_e$, uses a sliding window, $SW_e$, to handle the received uncertain IoT sensing data. Note that $|SW_e|\leq|SW|_{\max},\forall e$, where $|SW|_{\max}$ represents the predefined size constraint of all sliding windows. Each edge node, $N_e$, can directly communicate with the cloud server, $N_H$, where $e=1,2,\dots,m$. Each $N_e$ will continuously submit local collected IoT sensing data to $N_H$. For the cloud server, $N_H$, all reported information from every $N_e$ is treated as an input uncertain IoT data stream. $N_H$ continuously places the information received from the edge nodes into its sliding window $SW_H$ and $|SW_H|\leq|SW|_{\max}$. As stated in Definition~\ref{def:cb_sw}, both the cloud server and edge nodes employ count-based sliding windows.

\subsection{Problem Statement}
The objective of this study is to develop a time-efficient method to calculate and update the $k$-dominant skyline probability of all uncertain data items in the distributed edge-IoT environment. Although a low latency/computation time is regarded as a key performance indicator for future applications of multicriteria analysis on IoT data, the computation speed of the considered analytical application, $k$-dominant skyline, strongly relies on the dimensionality and volume of the input data.

Based on the above assumptions, this study proposes methods to solve the following two problems to improve the computation time for $k$-dominant skylines on uncertain IoT data: 
\begin{enumerate}
	\item How to efficiently update dominant probabilities of data items by using the dimensionality of data?
	\item How to effectively reduce the volume of input data?   
\end{enumerate}

\section{Proposed Distributed Indexing Schemes}
\label{sec:proposed_approach}
In this section, we introduce two proposed indexing schemes, MI and AI. 
Using these schemes, the proposed new theorem can determine an appropriate threshold to build sorted index tables to prune irrelevant data items. 
This design effectively reduced the amount of reference data required for dominant probability updates of the data items in response to the first problem statement. 
Furthermore, we applied the proposed indexing scheme to multiple edge-computing nodes and used this distributed edge framework to efficiently prune irrelevant data items. 
This approach effectively eliminated the irrelevant data and avoided unnecessary comparisons, thereby significantly improving the overall computation time. 
In summary, the combination of the indexing scheme and distributed edge framework was applied to address the second problem statement.



\subsection{Preprocessing}

\subsubsection{Data Normalization and Sorting}
The proposed MI method was designed to surpass the limitation of cross-dimensional comparisons. To achieve this goal, the first stage of MI, \emph{data normalization and dorting}, normalizes the values of each dimension to the same range and sorts the normalized items in ascending order. An example of a normalized and sorted dataset containing three data items is presented in Table~\ref{table:sorted_normalize_exmp}.

\subsubsection{Selection of Threshold Index Position}
\normalsize
	After the stage of data normalization and sorting, we obtained a normalized and sorted dataset, $SORTED(U)$, and each data item, $u\in SORTED(U)$, will be sorted in ascending order, denoted as $SORTED(u)$.
Thereafter, we employed a pointer $u_{\min}(k)$ to select an index position and used the value stored in this index position as a threshold $MI_{\min}(u,k)$. $u_{\min}(k)$ and $MI_{\min}(u,k)$ can be respectively expressed as
	\begin{align}
		u_{\min}(k) &\in \{0,1,...,k-1\}, \label{eq:4}\\
		MI_{\min}(u,k) &= SORTED(u)\left[u_{\min}(k)\right].\label{eq:5}
	\end{align}
Using~\eqref{eq:4} and~\eqref{eq:5}, we defined another pointer $u_{\max}(k)$ and one additional threshold $MI_{\max}(u,k)$ as 
	\begin{align}
		u_{\max}(k) &= u_{\min}(k) + (d-k), \label{eq:6}\\
		MI_{\max}(u,k) &= SORTED(u)\left[u_{\max}(k)\right]. \label{eq:7}
	\end{align}

Based on these two pointers and two thresholds, we proposed the following theorem:
\begin{thm}
	\label{thm:1}
	Given a $d$-dimensional space $S$, two data items $p,q\in S$, if $q_{\min}(k)=p_{\min}(k)$ and $MI_{\max}(p,k)<MI_{\min}(q,k)$, such that $q$ cannot $k$-dominate $p$.
\end{thm}
\begin{IEEEproof}
		Assuming that $q_{\min}(k)=p_{\min}(k)$, $MI_{\max}(p,k)<MI_{\min}(q,k)$ and $q$ $k$-dominates $p$, $q$ is not worse than $p$ in at least $k$ dimensions and $q$ outperforms $p$ in at least one dimension. 
	However, as $MI_{\max}(p,k)<MI_{\min}(q,k)$, $q$ has opportunities to outperform $p$ in the dimensions from index $0$ to index $(q_{\min}(k)-1)$ and from index $p_{\max}(k)$ to index $(d-1)$. In summary, $q$ contains at most $1+(q_{\min}(k)-1)+(d-1)-p_{\max}(k)=k-1$ opportunities to be not inferior or superior than $p$, indicating that $q$ may $(k-1)$-dominate $p$ at most. This contradicts the given assumption and the proof is established.
\end{IEEEproof}

\subsection{Scheme I: Middle Indexing (MI)}
\subsubsection{Construction of Sorted Middle Index Tables}
The proposed MI method sorts each item according to $MI_{\max}(u,k)$ and $MI_{\min}(u,k)$ to filter out the redundant items not required for comparison, reduce unnecessary calculations, and increase the speed of updates. A crucial consideration of this method is that the index position $u_{\min}(k)$ of each data item $u$ remain constant, i.e., all data items require the same baseline to be sorted.

MI includes two sorting strategies--according to the value of $MI_{\max}(u,k)$ and according to the value of $MI_{\min}(u,k)$. The purpose of constructing such an index table is to avoid recalculating and sorting the $MI_{\max}(u,k)$ and $MI_{\min}(u,k)$ of $u_{\text{sw}}$ every instant, which can accelerate the procedure of updating the probabilities of the $k$-dominant skyline items.

For each item flowing into a new stream, the proposed MI will use $MI_{\max}(u,k)$ as the key and insert it into the middle index table, $MIT_{\max}(k)$, in descending order. Similar to the index table, $MIT_{\max}(k)$, each new data item can be inserted into the index table $MIT_{\min}(k)$ in ascending order.

\begin{table}[!t]
	\caption{Example Of A Normalized and Sorted Dataset}
	\label{table:sorted_normalize_exmp}
	\centering	
	\begin{tabular}{|l|l|l|l|} 
		\hline
		Item & Index 0 & Index 1 & Index 2 \\
		\hline
		$u_1$ & 0 & 1 & 1 \\
		$u_2$ & 0 & 0.5 & 1 \\
		$u_3$ & 0 & 0.33 & 0.55  \\
		\hline
	\end{tabular}
\end{table}

\subsubsection{The Procedure of Updating $k$-Dominant Skyline Probability using Middle Indexing Scheme}
The procedural flowchart for updating $\mathbb{P}_{k\text{-sky}}(u_{\text{sw}})$ is illustrated in Fig.~\ref{fig:fig1}. The corresponding algorithm is described as a function $\mathsf{MI\_Update}()$ in Algorithm~\ref{alg:mi_update}. The operations from Lines~\ref{alg:mi_update:line1}--\ref{alg:mi_update:line9} implement the update of $\mathbb{P}_{k\text{-sky}}(u_{\text{sw}})$ in~\eqref{eq:obj_dominant_probability_update1} if the sliding window $SW$ is full. The update of $\mathbb{P}_{k\text{-sky}}(u_{\text{sw}})$ in~\eqref{eq:obj_dominant_probability_update2} is implemented from Lines~\ref{alg:mi_update:line10}--\ref{alg:mi_update:line16}. Moreover, lines~\ref{alg:mi_update:line3} and~\ref{alg:mi_update:line11} of Algorithm~\ref{alg:mi_update} use the proposed thresholds in~\eqref{eq:5} and~\eqref{eq:7} to review the conditions $MI_{\min}(u_{\text{old}},k)>MI_{\max}(u_{\text{sw}},k)$ and $MI_{\min}(u_{\text{new}},k)>MI_{\max}(u_{\text{sw}},k)$, respectively. If $MI_{\min}(u_{\text{old}},k)>MI_{\max}(u_{\text{sw}},k)$ holds at Line~\ref{alg:mi_update:line3}, $u_{\text{old}}$ cannot $k$-dominates any other items stored in following slots of sorted middle index table, $MIT_{\max}(k)$. Thus, the system does not execute any action because the following items in the sorted middle index table, $MIT_{\min}(k)$, fails to $k$-dominates $u_{\text{new}}$. Similarly, if $MI_{\min}(u_{\text{old}},k)>MI_{\max}(u_{\text{sw}},k)$ holds at Line~\ref{alg:mi_update:line11}, $u_{\text{new}}$ cannot $k$-dominate any items in $SW$ and the system does not execute any action. Otherwise, the system will review whether $u_{\text{new}}$ $k$-dominates $u_{\text{sw}}$. If yes, the system will update $\mathbb{P}_{k\text{-sky}}(u_{\text{sw}})$ using~\eqref{eq:obj_dominant_probability_update2} at Line~\ref{alg:mi_update:line14}.
The design of thresholds beneficially reduces numerous unnecessary comparisons. This benefit has been established in Theorem~\ref{thm:1}. 

After updating $\mathbb{P}_{k\text{-sky}}(u_{\text{sw}})$ of each remaining item $u_{\text{sw}}$, in the sliding window $SW$, the system calculated the $\mathbb{P}_{k\text{-sky}}(u_{\text{new}})$ of the new item $u_{\text{new}}$.
The detailed operations were described as the function $\mathsf{MI\_Calculate}()$ in Algorithm~\ref{alg:mi_cal_new}. Line~\ref{alg:mi_cal_new:line2} of Algorithm~\ref{alg:mi_cal_new} uses the proposed thresholds in~\eqref{eq:5} and~\eqref{eq:7} to review the condition $MI_{\max}(u_{\text{new}},k)<MI_{\min}(u_{\text{sw}},k)$. If $MI_{\max}(u_{\text{new}},k)<MI_{\min}(u_{\text{sw}},k)$ holds, $u_{\text{sw}}$ cannot $k$-dominate $u_{\text{new}}$. Thus, the system does nothing because the following items in the sorted middle index table $MIT_{\min}(k)$ cannot $k$-dominate $u_{\text{new}}$. If $MI_{\max}(u_{\text{new}},k)<MI_{\min}(u_{\text{sw}},k)$ does not hold, the system will review whether $u_{\text{sw}}$ $k$-dominates $u_{\text{new}}$. If yes, the system will calculate/update $\mathbb{P}_{k\text{-sky}}(u_{\text{new}})$ using~\eqref{eq:obj_dominant_probability} at Line~\ref{alg:mi_cal_new:line5}. 

Finally, the system will add the indexing information of $u_{\text{new}}$ to the proposed sorted middle index tables $MIT_{\max}(k)$ and $MIT_{\min}(k)$, which will assist in the future processing of $k$-dominant skylines. Such operations are described as function $\mathsf{MI\_Sort}()$ in Algorithm~\ref{alg:mi_sort}.


\begin{algorithm2e}[!t]
	\caption{$\mathsf{MI\_Update}()$}
	\label{alg:mi_update}
	\SetAlgoLined
	\KwIn{$SW$, $u_{\text{new}}$, $u_{\text{old}}$, $MIT_{\max}(k)$}	
	\KwOut{Updated $SW$}
	\If{$|SW|==|SW|_{\max}$}{\label{alg:mi_update:line1}
		\ForEach{$e$ in $MIT_{\max}(k)$}{
			\tcc{$e$ is $u_{\text{sw}}$}
			\uIf{$MI_{\min}(u_{\text{old}},k)>MI_{\max}(e,k)$}{\label{alg:mi_update:line3}
				\textbf{break}\;
			}
			\ElseIf{$u_{\text{old}}\prec_k e$}{
				$\mathbb{P}_{k\text{-sky}}(e)=		\mathbb{P}_{k\text{-sky}}(e)/(1-\mathbb{P}\left(u_{\text{old}})\right)$\;
			}
		}
	}\label{alg:mi_update:line9}
	\ForEach{$e$ in $MIT_{\max}(k)$}{\label{alg:mi_update:line10}
		\uIf{$MI_{\min}(u_{\text{new}},k)>MI_{\max}(e,k)$}{\label{alg:mi_update:line11}
			\textbf{break}\;
		}
		\ElseIf{$u_{\text{new}}\prec_k e$}{
			$\mathbb{P}_{k\text{-sky}}(e)=		\mathbb{P}_{k\text{-sky}}(e)\times(1-\mathbb{P}\left(u_{\text{new}})\right)$\label{alg:mi_update:line14}\;
		}
	}\label{alg:mi_update:line16}
	\Return $SW$\;
\end{algorithm2e}
%

\begin{algorithm2e}[!t]
	\caption{$\mathsf{MI\_Calculate}()$}
	\label{alg:mi_cal_new}
	\SetAlgoLined
	\KwIn{$SW$, $u_{\text{new}}$, $MIT_{\min}(k)$}
	\KwOut{$\mathbb{P}_{k\text{-sky}}(u_{\text{new}})$}
	\ForEach{$e$ in $MIT_{\min}(k)$}{
		\uIf{$MI_{\max}(u_{\text{new}},k)<MI_{\min}(e,k)$}{\label{alg:mi_cal_new:line2}
			\textbf{break}\;
		}
		\ElseIf{$e\prec_k u_{\text{new}}$}{
			$\mathbb{P}_{k\text{-sky}}(u_{\text{new}})=		\mathbb{P}_{k\text{-sky}}(u_{\text{new}})\times(1-\mathbb{P}\left(e)\right)$\label{alg:mi_cal_new:line5}\;
		}
	}
	\Return $\mathbb{P}_{k\text{-sky}}(u_{\text{new}})$\;
\end{algorithm2e}
%

\begin{algorithm2e}[!t]
	\caption{$\mathsf{MI\_Sort}()$}
	\label{alg:mi_sort}
	\SetAlgoLined
	\KwIn{$u_{\text{new}}$}
	\KwOut{$MIT_{\max}(k)$, $MIT_{\min}(k)$}		
	insert $u_{\text{new}}$ to $MIT_{\max}(k)$\;
	insert $u_{\text{new}}$ to $MIT_{\min}(k)$\;
	\Return $MIT_{\max}(k)$, $MIT_{\min}(k)$\;
\end{algorithm2e}

\subsubsection{Running Example of MI}
\begin{figure}[!t]
	\centering
	\includegraphics[width=.8\columnwidth]{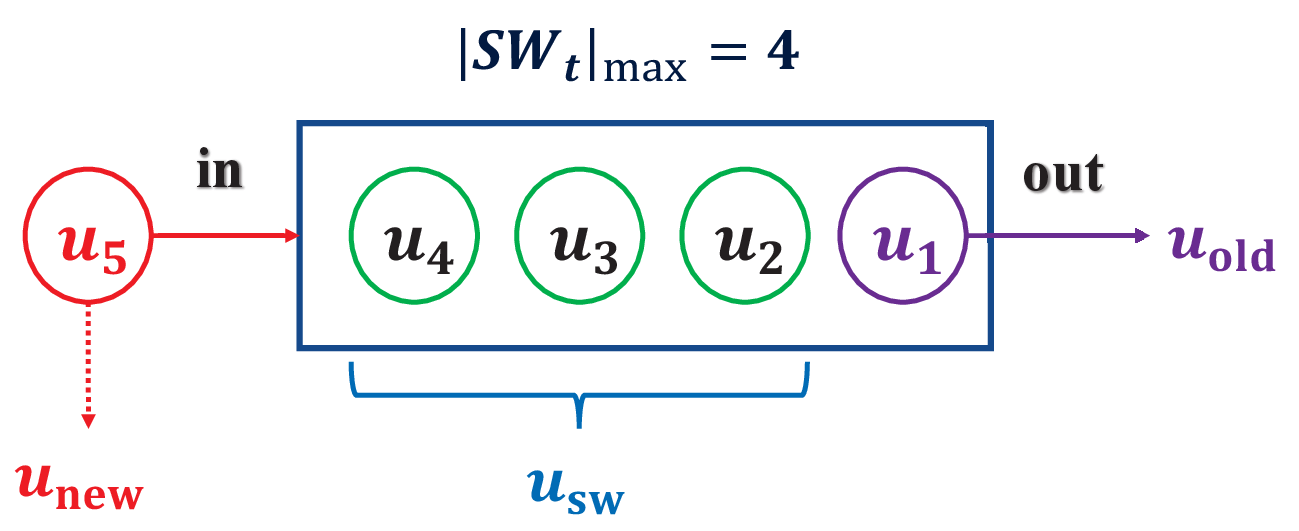}
	\caption{Scenario of a running example, where $|SW|_{\max}=4$, $u_1$ is going to leave the sliding window, and $u_5$ is the new data item $u_{\text{new}}$ to the sliding window at $t=5$}
	\label{fig:mi:running_ex}
\end{figure}

Consider a scenario of a running example in Fig.~\ref{fig:mi:running_ex} and the corresponding dataset in Table~\ref{tab:mi:running_ex}. If $|SW|_{\max}=4$, $k=3$, and the data items $u_1, u_2,\dots, u_5$ enter the sliding window sequentially, then $SW_4=\{u_1,u_2,u_3,u_4\}$. If $t=5$, $SW_4=\{u_2,u_3,u_4,u_5\}$, $u_5$ represents the new data and $u_1$ becomes outdated. If we select index 2 and index 3 as the thresholds $u_{\min}(3)$ and $u_{\max}(3)$, respectively, the sorted middle index tables $MIT_{\max}(3)$ and $MIT_{\min}(3)$ will be constructed as Tables~\ref{tab:mi:MIT_max_ex} and~\ref{tab:mi:MIT_min_ex}, respectively, for $t=5$.

Since $u_5$ becomes outdated at $t=5$ and will be remove from the sliding window $SW$, the 3-dominant probabilities of the items $u_{\text{sw}}$ remaining in $SW$ may vary. 
The system will use $MI_{\min}(u_1,3)=70$ in Table~\ref{tab:mi:running_ex} to compare with $MI_{\max}(u,3)$ in $MIT_{\max}(3)$ (Table~\ref{tab:mi:MIT_max_ex}). 
If $MI_{\max}(u,3)<MI_{\min}(u_1,3)$, the system will not continue to compare with the items stored after $u_1$. As $u_1$ cannot dominate the items stored after $u_1$, the system does not need to update the $3$-dominant probabilities of these items. 
In this running example, $MI_{\max}(u_4,3)=60<MI_{\min}(u_1,3)=70$, such that the 3-dominant probabilities of the items stored after $u_4$ do not require to be inspected and updated. 
Conversely, during the influx of a new item $u_{\text{new}}=u_5$, the system will review and update the three dominant probabilities of the items remaining in the $SW$ similarly.

Furthermore, the system will determine the objects remaining in the $SW$ that can 3-dominate the new item $u_{\text{new}}=u_5$ by comparing $MI_{\max}(u_5,3)=80$ (in Table~\ref{tab:mi:running_ex}) with $MIT_{\min}(3)$ in $MI_{\min}(u,3)$. 
If $MI_{\max}(u_5,3)<MI_{\min}(u,3)$, the items stored after $u$ cannot 3-dominate $u_5$. 
For the considered running example, we observed that $MI_{max}(u_5,3)=80<MI_{min}(u_2,3)=90$ and the system does not need to review and update the 3-dominant probabilities of the items stored after $u_2$.

Finally, the system will insert $u_5$ to $MIT_{\min}(3)$, $MIT_{\max}(3)$, and $SW$. At this instant, the entry and departure operations of certain items within a certain period have all been completed.

\begin{table}[!t]
	\caption{Corresponding Dataset of The Running Example}
	\label{tab:mi:running_ex}
	\centering	
	\begin{tabular}{|l|l|l|l|l|l|}    
		\hline
		Item & Index 0 & Index 1 & Index 2 & Index 3 \\
		\hline
		$u_1$ & 30 & 40 & 70 & 70 \\
		$u_2$ & 50 & 80 & 90 & 90 \\
		$u_3$ & 20 & 40 & 40 & 90 \\
		$u_4$ & 20 & 30 & 50 & 60 \\
		$u_5$ & 40 & 60 & 80 & 80 \\
		\hline
	\end{tabular}
\end{table}

\begin{table}[!t]
	\caption{Example of $MIT_{max}(3)$}
	\label{tab:mi:MIT_max_ex}
	\centering	
	\begin{tabular}{|l|l|}    
		\hline
		Item & $MI_{\max}(u,3)$ \\
		\hline
		$u_2$ & 90 \\
		$u_3$ & 90 \\
		$u_4$ & \textbf{60} \\
		\hline
	\end{tabular}
\end{table}

\begin{table}[!t]
	\caption{Example of $MIT_{min}(3)$}
	\label{tab:mi:MIT_min_ex}
	\centering	
	\begin{tabular}{|l|l|}    
		\hline
		Item & $MI_{\min}(u,3)$ \\
		\hline
		$u_3$ & 40 \\
		$u_4$ & 50 \\
		$u_2$ & \textbf{90} \\
		\hline
	\end{tabular}
\end{table}

\subsection{Scheme II: All Indexing (AI)}
In the MI scheme, the system utilized precomputed threshold index positions $u_{\min}$ and $u_{\max}$ to construct the sorted middle index tables $MIT_{\max}(k)$ and $MIT_{\min}(k)$ for the given $k$. 
Using $MIT_{\max}(k)$ and $MIT_{\min}(k)$, the system can determine two thresholds $MI_{\max}(u,k)$ and $MI_{\min}(u,k)$ to prune irrelevant data items, and consequently, improve the efficiency of the $k$-dominant skyline probability update. 
For reviewing the dominance between varying items under the same baseline, the system needs to sort $MIT_{\max}(k)$ and $MIT_{\min}(k)$ with the fixed threshold index positions, $u_{\min}$ and $u_{\max}$. 
This finding signifies that $u_{\min}$ and $u_{\max}$ cannot vary dynamically, which undermines the utility of Theorem~\ref{thm:1}. 
Consequently, certain irrelevant data items will not be filtered out. 
For instance, when new data items arrive, they need to be sorted according to the $u_{\min}$ and $u_{\max}$ obtained in the preprocessing stage. 
Otherwise, all data items cannot be sorted and compared with the same baseline.
Therefore, we proposed the AI scheme to more effectively filter based on Theorem~\ref{thm:1}. 


\subsubsection{The Procedure of Updating $k$-Dominant Skyline Probability using All Indexing Scheme}
Unlike the MI scheme, the AI scheme does not construct and use sorted MI tables to update the $k$-dominant skyline probabilities of $u_{\text{sw}}$ and $u_{\text{new}}$ ($\mathbb{P}_{k\text{-sky}}(u_{\text{sw}})$ and $\mathbb{P}_{k\text{-sky}}(u_{\text{new}})$). When a new data item $u_{\text{new}}$ enters the system, the AI scheme will update the $k$-dominant skyline probabilities of $u_{\text{sw}}$ and $u_{\text{new}}$ by directly using~\eqref{eq:4}--\eqref{eq:7} and Theorem~\ref{thm:1}. Thus, the AI needs to compute all possible $u_{\max}(k)$ and $u_{\min}(k)$ along with the corresponding possible $MI_{\max}(u,k)$ and $MI_{\min}(u,k)$. Thereafter, the AI scheme uses Theorem~\ref{thm:1} to filter out irrelevant data items, which reduces the computation cost of updating the $k$-dominant skyline probabilities of the remaining items $u_{\text{sw}}$ in $SW$ and the new item $u_{\text{new}}$. When an old data item $u_{\text{old}}$ exits the sliding window, the AI scheme uses the same method to compute $\mathbb{P}_{k\text{-sky}}(u_{\text{sw}})$ and $\mathbb{P}_{k\text{-sky}}(u_{\text{new}})$.

\begin{algorithm2e}[!t]
	\caption{$\mathsf{AI\_Update}()$}\label{algo:ai:updateSW}
	\KwIn{$SW,u_{\text{new}},u_{\text{old}},AIT_{\max}$\;}
	\KwOut{Updated $SW$\;}
	$AIT_{\max}\leftarrow \mathsf{Cal\_AIT\_Max}()$\;
	\If{$|SW|==|SW|_{\max}$}{
		\ForEach{$e$ in $AIT_{\max}$}{
			\tcc{$e$ is $u_{\text{sw}}$}
			\If{$u_{\text{old}} \prec_k e$}{
				$\mathbb{P}_{k\text{-sky}}(e) = \mathbb{P}_{k\text{-sky}}(e) / (1-\mathbb{P}(u_{\text{old}}))$\;
			}
		}
	}
	$AIT_{\max}\leftarrow \mathsf{Cal\_AIT\_Max}()$\;
	\ForEach{$e$ in $AIT_{\max}$}{
		\If{$u_{\text{new}} \prec_k e$}{
			$\mathbb{P}_{k\text{-sky}}(e) = \mathbb{P}_{k\text{-sky}}(e) \times (1-\mathbb{P}(u_{\text{new}}))$\;
		}
	}
	\Return $SW$\;
\end{algorithm2e}

\begin{algorithm2e}[!t]
	\caption{$\mathsf{Cal\_AIT\_Max}()$}\label{algo:AITMaxsort}
	\KwIn{$u_{\text{new}}$ or $(u_{\text{old}}), u_{\text{sw}}$\;}
	\KwOut{$AIT_{\max}$\;}
	$u_{\text{temp}}\leftarrow u_{\text{new}}$ (or $u_{\text{old}}$)\;
	initialize $AIT_{\max}$ as an empty list\;
	\ForEach{$e$ in $SW$}{
		\If{$MI_{\min}(u_{\text{temp}},k)\leq MI_{\max}(e,k)$}{
			append $e$ to $AIT_{\max}$ \;
		}
	}
	\Return $AIT_{\max}$ \;
\end{algorithm2e}

The procedure of updating $\mathbb{P}_{k\text{-sky}}(u_{\text{sw}})$ using the AI scheme is described below. When a new data item $u_{\text{new}}$ enters the system, if the sliding window is full ($|SW(t)|=|SW|_{\max}$), the system with the AI scheme firstly uses indexed values $MI_{\max}(u_{\text{sw}},k)$, $MI_{\min}(u_{\text{old}},k)$ and Theorem~\ref{thm:1} to rapidly filter out the data items that cannot be possibly $k$-dominated by $u_{\text{old}}$. After precisely inspecting whether the remaining data items can be $k$-dominated by $u_{\text{old}}$, the system derives a temporary set, expressed as follows:
\begin{equation}
	U_{\text{sw}}'=\{u|\forall u \in U_{\text{sw}}, u_{\text{old}} \prec_k u\}.
\end{equation}
If $U_{\text{sw}}'\neq \emptyset$, the system with the AI scheme will employ~\eqref{eq:6} to update $\mathbb{P}_{k\text{-sky}}(u), \forall u\in U_{\text{sw}}'\subseteq U_{\text{sw}}$. 

Second, the system requires to update the data items in $u_{\text{sw}}$ that are $k$-dominated by $u_{\text{new}}$. In this step, similar to the preceding step, the system uses the indexed values $MI_{\max}(u_{\text{sw}},k)$, $MI_{\min}(u_{\text{new}},k)$ and Theorem~\ref{thm:1} to filter out the data items that cannot be possibly $k$-dominated by $u_{\text{new}}$. After precisely reviewing whether the remaining data items can be $k$-dominated by $u_{\text{new}}$, the system obtains a temporary set:
\begin{equation}
	U_{\text{sw}}'=\{u|\forall u \in U_{\text{sw}}, u_{\text{new}} \prec_k u\}.
\end{equation}
If $U_{\text{sw}}'\neq \emptyset$, the system uses~\eqref{eq:7} to update $\mathbb{P}_{k\text{-sky}}(u), \forall u\in U_{\text{sw}}'\subseteq U_{\text{sw}}$. For the procedure conducted in the above two steps, please refer to Algorithms~\ref{algo:ai:updateSW} and~\ref{algo:AITMaxsort}.

\begin{algorithm2e}[!t]
	\caption{$\mathsf{AI\_Calculate}()$}\label{algo:ai:updateNew}
	\KwIn{$u_{\text{new}},AIT_{\min}$\;}
	\KwOut{$\mathbb{P}_{k\text{-sky}}(u_{\text{new}})$\;}
	$AIT_{\min}\leftarrow \mathsf{Cal\_AIT\_Min}()$\;
	\ForEach{$e$ in $AIT_{\min}$}{		
		\If{$e \prec_k u_{\text{new}}$}{
			$\mathbb{P}_{k\text{-sky}}(u_{\text{new}}) = \mathbb{P}_{k\text{-sky}}(u_{\text{new}}) \times (1-\mathbb{P}(e))$\;
		}		
	}
	\Return $\mathbb{P}_{k\text{-sky}}(u_{\text{new}})$\;
\end{algorithm2e}

\begin{algorithm2e}[!t]
	\caption{$\mathsf{Cal\_AIT\_Min}()$}\label{algo:AITMinsort}
	\KwIn{$u_{\text{new}}, u_{\text{sw}}$\;}
	\KwOut{$AIT_{\min}$\;}
	initialize $AIT_{\min}$ as an empty list\;
	\ForEach{$e$ in $SW$}{
		\If{$MI_{\min}(e,k)\leq MI_{\max}(u_{\text{new}},k)$}{
			append $e$ to $AIT_{\min}$\;
		}
	}
	\Return $AIT_{\min}$ \;
\end{algorithm2e}

Thereafter, in the similar manner, the system uses all the indexed values $MI_{\max}(u_{\text{sw}},k)$, $MI_{\min}(u_{\text{new}},k)$, and Theorem~\ref{thm:1} to prune those data items that cannot possibly $k$-dominate $u_{\text{new}}$. After the system inspects whether the remaining data items are $k$-dominated by $u_{\text{new}}$, the system derives a temporary set
\begin{equation}
	U_{\text{sw}}'=\{u|\forall u \in U_{\text{sw}}, u \prec_k u_{\text{new}}\}.
\end{equation}
If $U_{\text{sw}}'\neq \emptyset$, the system uses~\eqref{eq:7} with the data items in $U_{\text{sw}}'$ to derive $\mathbb{P}_{k\text{-sky}}(u_{\text{new}})$. The procedure for this step is presented in Algorithms~\ref{algo:ai:updateNew} and~\ref{algo:AITMinsort}.

\subsubsection{Running Example of AI}
The AI scheme uses all possible index positions to obtain the indexed threshold values for pruning data items. 
Considering the same example in Fig.~\ref{fig:mi:running_ex}, AI will use the following three combinations of indexes: 1) $u_{\min}(3)=0,u_{\max}(3)=1$; 2) $u_{\min}(3)=1,u_{\max}(3)=2$; and 3) $u_{\min}(3)=2,u_{\max}(3)=3$ as filtering conditions. 
First, as $u_{\text{old}}=u_1$ exists, the system will determine the data items in the sliding windows that are altering their own $k$-dominant probabilities, i.e., $U_{\text{sw}}'=\{u|\forall u \in U_{\text{sw}}, u_1 \prec_k u\}$, where $U_{\text{sw}}=\{u_2,u_3,u_4\}$. 
For each data item $u$ in $U_{\text{sw}}$, the system uses all indexed threshold values corresponding to the above three combinations of indices to prune the data items. 
For each combination if $MI_{\min}(u_1,k)>MI_{\max}(u,k),\forall u\in U_{\text{sw}}$, $u$ will be filtered out. 
In this running example of $k=3$, $u_2$ and $u_3$ does not satisfy all the above-mentioned three combinations of filtering conditions such that they are not filtered out at this step. 
For $u_{\min}(3)=2,u_{\max}(3)=3$, $MI_{\min}(u_1,3)=70>MI_{\max}(u_4,3)=60$, so $u_4$ will be pruned out. 
Until this step, by leveraging the proposed AI scheme, the system can efficiently obtain a temporary set $U_{\text{sw}}'=\{u_2,u_3\}$. 
This temporary set includes data items that may need to update their $3$-dominant probabilities.  
However, the system still needs to review whether $u_1$ can assuredly $3$-dominates $u_2$ and $u_3$. 
If yes, the system will update the $3$-dominant probability of each data item that is exactly dominated by $u_{\text{old}}=u_1$. 
For the other two scenarios: 1) the $3$-dominant probability of each data item $u\in U_{\text{sw}}$ is updated, where $u_1\prec_3 u$ and 2) the $3$-dominant probability of $u_{\text{new}}=u_5$ is updated, based on which the system uses the aforementioned three combinations of the indices (filtering conditions) to prune irrelevant data items in the same manner.

\subsection{Complexity Discussion}
After introducing the proposed MI and AI schemes, we discuss their time complexity in this section. As discussed earlier, $d$ denotes the data dimension and $|SW|_{\max}$ represents the size constraint of the sliding window on a server or edge node. In the first step of the preprocessing phase (i.e., data normalization and sorting), a normalized and sorted dataset should be constructed in both MI and AI. Using a linear time sorting algorithm (e.g. bucket sort), the computation cost of this step is $O(|SW|_{\max}\cdot d)$. 

In the second step of the preprocessing phase, the threshold index positions and thresholds are selected according to~\eqref{eq:4}--\eqref{eq:7}. 
Therefore, this operation requires a computation time of only $O(1)$. 
Subsequently, the system enters the second phase of updating the $k$-dominant skyline probability. 
In this phase, MI uses two middle index tables, $MIT_{\max}(k)$ and $MIT_{\min}(k)$, and requires a computation time of $O(|SW|_{\max})$ to update the dominant probabilities of the relevant data items in $SW$. 
Conversely, the AI uses all combination of $(u_{\min}(k),u_{\max}(k))\in\{(0,d-k),(1,d-k+1),\dots,(k-1,d)\}$ to filter out irrelevant data items and update dominant probabilities of the relevant data items in $SW$, thereby requiring $O(|SW|_{\max}\cdot k)$ for computation. 
Therefore, the time complexity of the MI and AI schemes in the worst scenario will be $O(|SW|_{\max}\cdot (d+1))$ and $O(|SW|_{\max}\cdot (d+k))$, respectively. 
In fact, both MI and AI schemes bear the same time complexity, $O(|SW|_{\max}\cdot d)$, because $k\leq d$.

However, the occurrence of the worst case is highly infrequent. Owing to effective data pruning, the number of reference data items for updating the dominant probability in the second phase is typically less than $|SW|_{\max}$. Suppose the average numbers of the data items referenced for each update of dominant probability in the MI and AI schemes are $\bar{r}_{\rm MI}$ and $\bar{r}_{\rm AI}$, respectively, where $\bar{r}_{\rm MI}\leq |SW|_{\max}$ and $\bar{r}_{\rm AI}\leq |SW|_{\max}$. Accordingly, the time complexity of the MI and AI schemes in the median occurrence case will be $O(|SW|_{\max}\cdot d+\bar{r}_{\rm MI})$ and $O(|SW|_{\max}\cdot d+\bar{r}_{\rm AI}\cdot k)$, respectively.


\section{Simulation Results}
\label{sec:simulation}
In this section, we conducted several simulations to verify the performance of the proposed MI and AI schemes, including their comparison with the existing method PKDS-CI~\cite{li2019parallel} under two distinct scenarios/environments. In Scenario I, only a single computing node is employed to simulate a centralized computing environment, indicating that the master node is a poker node. In Scenario II, five computing/worker nodes are used including a master node to simulate a distributed edge-computing environment. The master node is responsible for collecting distributed computing results and deriving the global $k$-dominant skyline.

As the objective of this research is to improve the computation time (or latency) of the $k$-dominant skyline query (problem statement described in Section~\ref{sec:problem}), we analyzed the performance of the proposed MI and AI schemes and compared it with that of the PKDS-CI method in terms of the average computation time.
Recall the problem statement described in Section~\ref{sec:problem}, the objective of this work is to improve the computation time (or latency) of the $k$-dominant skyline query. Hence, we will discuss the performance of our proposed MI and AI schemes and the compared PKDS-CI method in terms of average computation time. In particular, we investigated the influence of the two major factors on the average computation time of each method, i.e., $k$ value and the sliding window size $|SW|_{\max}$ in Scenarios I and II.

\subsection{Scenario I: Centralized Computing Environment}
In scenario I, we measured the performance of each scheme on a centralized computing node. The simulations were performed on a computer with an Intel Core i5-4460 CPU, 16 GB DDR3 RAM, and Windows 10. In total, six virtual machines (5 worker nodes and 1 master node) were deployed using Ubuntu 16.04. The simulation of Scenario I was implemented in Python 3.7.4 environment. The output data were the average of 30 iterations/results. The settings of the simulation parameters for Scenario I are presented in Table~\ref{table:simulation_setting_1}.

\begin{table}[!t]
	\centering
	\caption{Parameter Settings for Scenario I}
	\label{table:simulation_setting_1} 
	\begin{tabular}{l|l|l}
		\hline
		\textbf{Parameter} & \textbf{Values} & \textbf{Default Value}\\ \hline 
		Data dimensionality & 12 & 12 \\ 
		$k$ & 7, 8, 9, 10, 11 & 11 \\ 
		$|SW|_{\max}$ & 100, 200, \dots, 1,000 & 500 \\ 
		The number of data items & 10,000 & 10,000 \\ 
		\hline
	\end{tabular}
\end{table}

\begin{figure}[!t]
	\centering
	\includegraphics[width=\columnwidth]{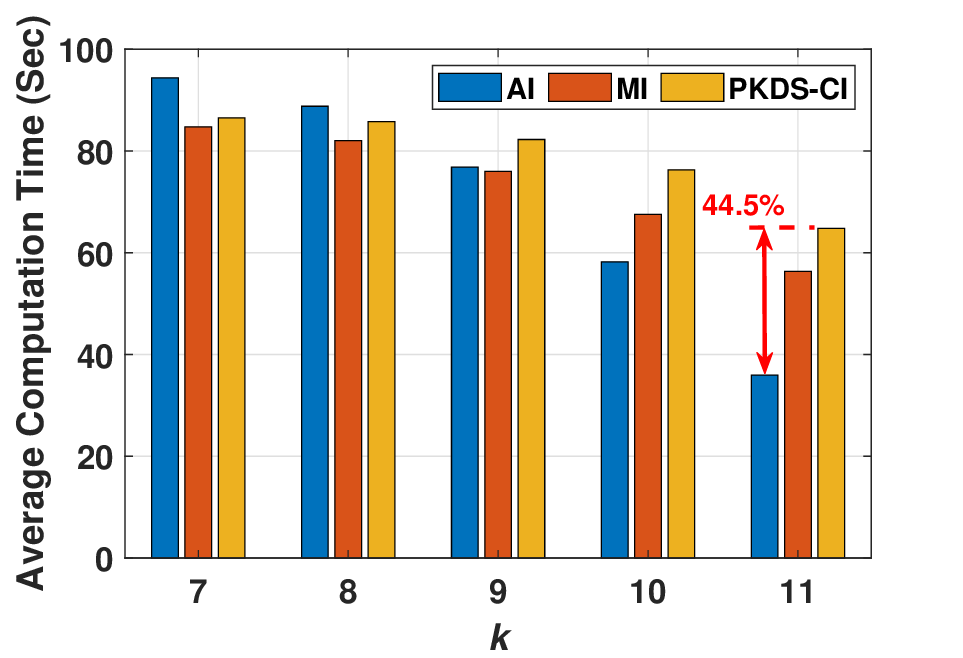}
	\caption{Computation time of various indexing schemes with varying $k$ values on a centralized computing node}
	\label{fig:query_k_value_centralized}
\end{figure}

First, We discuss the impact of $k$ value on the computation time. According to the results illustrated in Fig.~\ref{fig:query_k_value_centralized}, we determined that when the $k$-value was larger and approached the value of the data dimensions $d$, the computation cost of MI and PKDS-CI was less.
This is because for a small $k$, an item is less likely to be dominated by another item. 
Regardless of using any indexing scheme for data pruning, the effect of data pruning worsened as $k$ decreased. 
Notably, the proposed MI scheme outperformed PKDS-CI by about 2\% to 13\% for k varying from 7 to 11. 
Conversely, the proposed AI scheme outperformed PKDS-C by about 7\% to 44.5\% for $k$ varying from 9 to 11. 
If $k\leq 8$, the AI scheme delivered a worse performance than PKDS-C, as AI inspects all the combinations of the index positions, $u_{\max}(k)$ and $u_{\min}(k)$, for pruning data. 
Furthermore, if $k$ decreases, a data item cannot easily $k$-dominate another data item. 
This phenomenon increases the computation time of the AI scheme as it compares the attribute values between the data items but cannot effectively prune the data.
Recall that the average time complexity of AI is $O(|SW|_{\max}\cdot d+\bar{r}_{\rm AI}\cdot k)$ and $\bar{r}_{\rm AI}$ is typically much larger than $k$. 
These simulation results verify that the pruning effect is poor for small values of $k$ and $\bar{r}_{\rm AI}$ remains exceedingly large after pruning.
If $k$ increases, a data item can more easily $k$-dominate another data item. 
Thus, in such cases, the AI scheme delivered superior performance in terms of computation time, as it utilized a greater number of combinations of index positions ($u_{\max}(k)$ and $u_{\min}(k)$) than MI and PKDS-CI for data pruning, which significantly reduced $\bar{r}_{\rm AI}$. 
Therefore, the proposed AI scheme can more effectively prune data. 
In summary, the AI solution is more suitable for situations  with relatively large values of $d$ and $k$.

%
\begin{figure}[!t]
	\centering
	\includegraphics[width=\columnwidth]{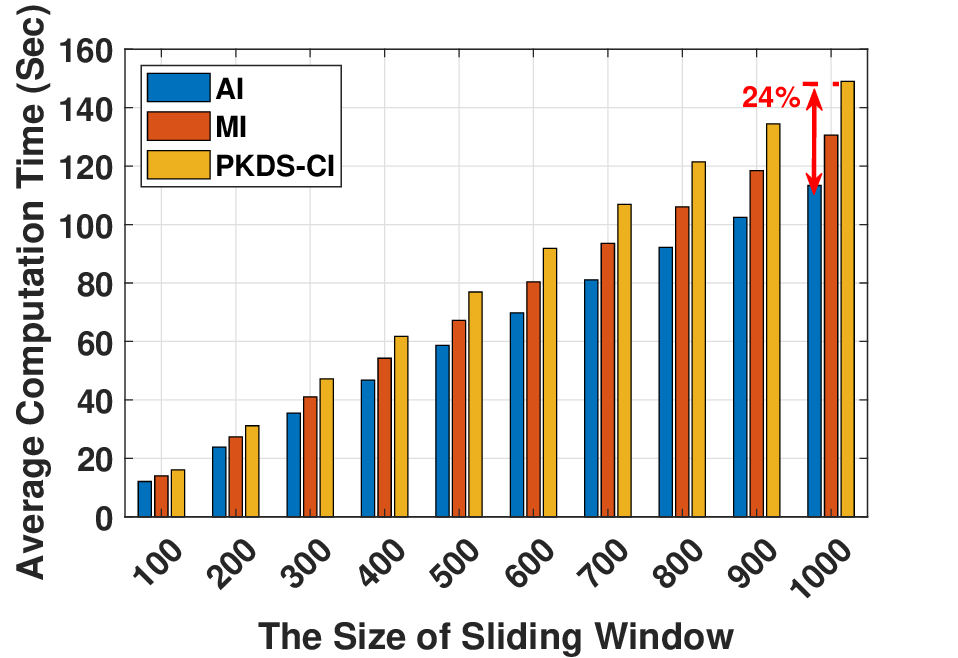}
	\caption{Computation time of various indexing schemes using varying size constraints of sliding windows, $|SW|_{\max}$, on a centralized computing node}
	\label{fig:SW_size_centralized}
\end{figure}

Additionally, we determined the influence of the size constraint of the sliding windows $|SW|_{\max}$ in the simulation of scenario I. 
As depicted in Fig.~\ref{fig:SW_size_centralized}, the computation cost of all the compared schemes increased with $|SW|_{\max}$ from $100$ to $1000$, because every update or evaluation of the $k$-dominant skyline probability of $u_{\text{new}}$ compared $u_{\text{new}}$ with all items in the sliding window for filtering out data that are $k$-dominated. 
As expected, the computation time increased when $SW$ contained a greater number of items. The results indicated that the extent to which MI and AI can outperform PKDS is not affected by the sliding window size. As portrayed in Fig.~\ref{fig:SW_size_centralized}, the proposed MI and AI schemes decreased the required computation time by about 12\% and 24\% compared to PKDS, respectively.

\subsection{Scenario II: Distributed Edge Computing Environment}
In Scenario II, we measured the performance of each scheme in a simulated distributed edge computing environment. We execute this simulation on a computer with an Intel Core i7-9700 CPU, 64 GB DDR4 RAM, and Windows 10. We deployed six virtual machines (5 worker nodes and 1 master node) using Ubuntu 16.04 with Apache Spark 2.4.4 platform~\cite{ZahariaXinEtAl16cacm}. The simulations were implemented in Python 3.7.4 environment. The average of 10 iterations/results from the simulations was considered the output data. The settings of the simulation parameters for Scenario II are presented in Table~\ref{table:simulation_setting_2}.

\begin{figure}[!t]
	\centering
	\includegraphics[width=\columnwidth]{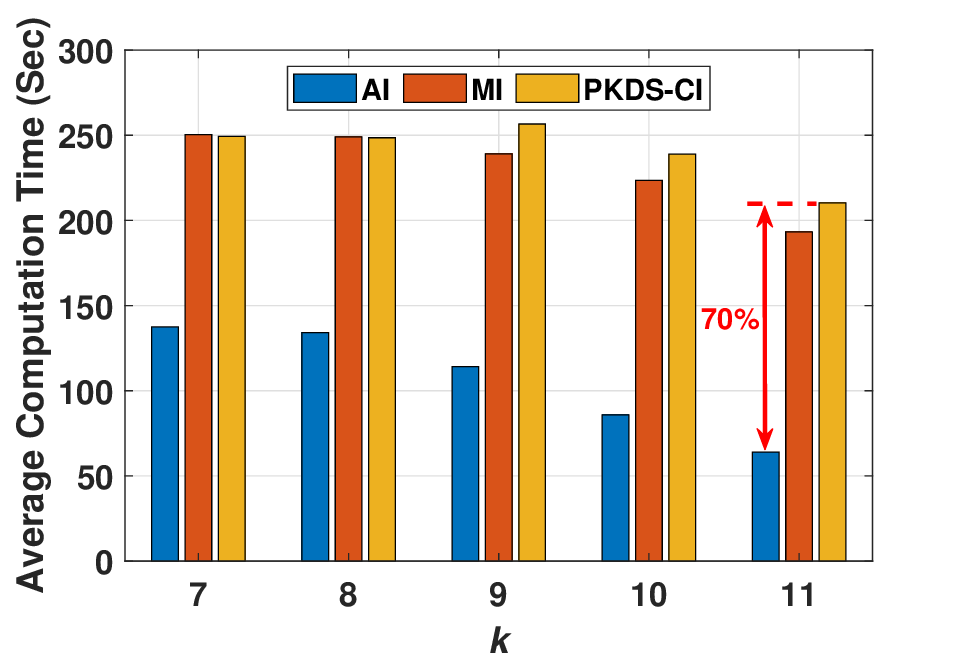}
	\caption{Computation time of various indexing schemes for varying $k$ values on distributed edge computing nodes}
	\label{fig:query_k_value_edge}
\end{figure}

\begin{figure}[!t]
	\centering
	\includegraphics[width=\columnwidth]{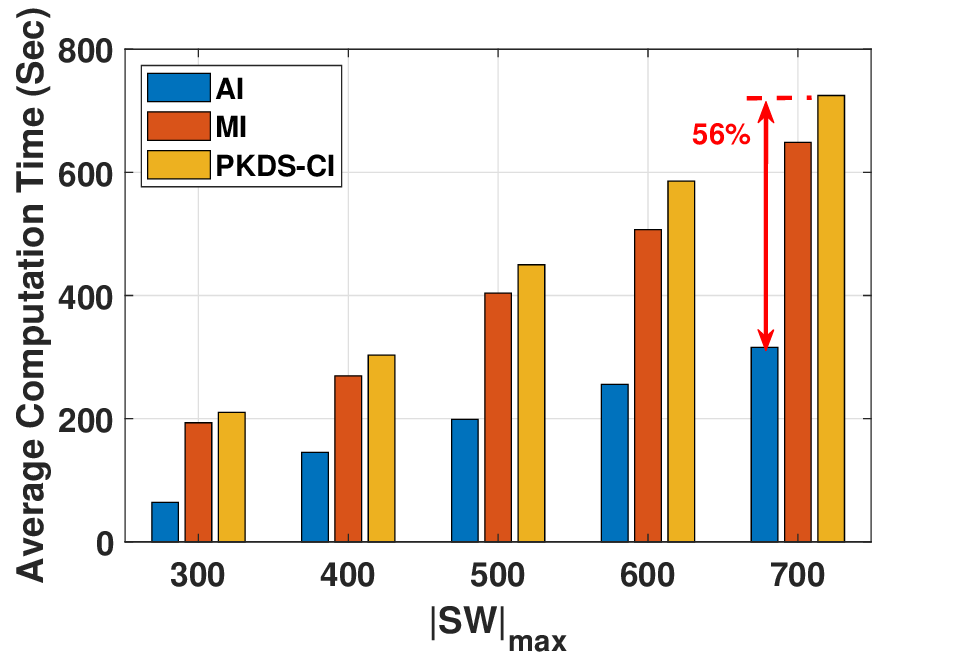}
	\caption{Computation time of various indexing schemes while using varying size constraints of sliding windows, $|SW|_{\max}$, on distributed edge computing nodes}
	\label{fig:SW_size_edge}
\end{figure}

First, we discuss the impact of $k$ value on computation time. The results obtained with various $k$ values from 7 to 11 is depicted in Fig.~\ref{fig:query_k_value_edge}. 
As $k$ increases, the performance of AI, MI, and PKDS-CI in terms of the computation time decreases. As $k$ increases, the AI and MI outperformed PKDS-CI by about 70\% and 8.06\%, respectively. 
Ultimately, we performed simulations to quantify the impact of the sliding window size on the computation time. The results are presented in Fig.~\ref{fig:SW_size_edge}. As observed from the results, the computation time increased with the sliding window size.  If the size of sliding window ranged from 300 to 700, MI was about 8\% to 13\% faster than PKDS-CI and AI was about 52\% to 56\% faster than PKDS-CI.  This result demonstrates that the lead does not significantly vary with the size of the sliding window.

\begin{table}[!t]
	\centering
	\caption{Parameter Settings for Scenario II}
	\label{table:simulation_setting_2} 
	\begin{tabular}{l|l|l}
		\hline
		\textbf{Parameter} & \textbf{Values} & \textbf{Default Value}\\ \hline 
		Data dimensionality & 12 & 12 \\ 
		$k$ & 7, 8, 9, 10, 11 & 11 \\ 
		$|SW|_{\max}$ & 300, 400, 500, 600, 700 & 300 \\ 
		The number of data items & 10,000 & 10,000 \\ 
		\hline
	\end{tabular}
\end{table}

\subsection{Comparison Summary}
As listed in Table~\ref{table:summary_of_comparisons}, the comparison of three schemes varied across the six distinct scenarios. The comparison characteristics were classified in three categories: best, medium, and worst.



According to the present simulation results, the MI scheme displayed medium computational performance across various simulation scenarios in a centralized environment. For large values of $d$, $k$ and $|SW|$, the AI scheme delivered the best computational performance. Overall, the conventional method PKDS-CI delivered the worst performance. The AI scheme delivered the worst performance only when the difference between $d$ and $k$ becomes exceedingly large. In such scenarios, the AI scheme cannot effectively prune the irrelevant data items, thereby conducting unnecessary dominance inspections and probability updates. Furthermore, the present simulation results revealed that the computation time for all comparison schemes were proportional to the size of the sliding window. Among these schemes, the AI scheme exhibited the flattest trend of computation time growth.

In a distributed edge-computing environment, AI consistently outperformed other comparisons in all considered scenarios. In all test scenarios, MI exhibited medium/moderate computational performance and the comparison method, PKDS-CI, was outperformed in all test scenarios. Only in the case of large $d$ and small $k$, the computational performances of PKDS-CI and MI were at par.

\begin{table}[!t]
	\centering
	\caption{Comparative Summary of Simulation Performance}
	\label{table:summary_of_comparisons}
	\begin{tabular}{|l|l|lll|}
		\hline
		\multirow{2}{*}{Model} & \multirow{2}{*}{Scenario} &\multicolumn{3}{c|}{Method} \\ \cline{3-5}
		& & MI & AI & PKDS-CI \\ \hline
		\multirow{3}{*}{Centralized}  &  Large $d$ and $k$  &  Medium &  Best  & Worst \\
		&   Large $d$ and small $k$  &  Medium &  Worst  & Best \\
		&   Large $|SW|$  &   Medium    &   Best     &   Worst       \\  \hline
		\multirow{3}{*}{Distributed}  &  Large $d$ and $k$  &  Medium  &    Best  &  Worst   \\
		&   Large $d$ and small $k$  &  Medium &    Best  & Medium \\
		&   Large $|SW|$  &    Medium    &      Best     &   Worst       \\ \hline
	\end{tabular}
\end{table}


\section{Conclusion}
\label{sec:conclusion}
As evaluating the $k$-dominant skyline probability of each data item in an uncertain data stream requires an enormous amount of computation, the theorem proposed in this study can effectively and rapidly determine the $k$-dominant relationship between two items. We applied this theorem for the derivation of the $k$-dominant skyline. In addition, we proposed two highly efficient indexing schemes, MI and AI, to effectively filter out several items that did not require any comparison, which significantly accelerated the calculation/update speed. Furthermore, we applied the proposed schemes to a simulated distributed edge-computing environment and conducted certain simulations to measure their performance. According to the simulation results, the distributed MI and AI decreased the computation time by about 13\% and 56\% compared with the existing method. In particular, for processing high-dimensional uncertain data, the AI scheme can outperform the existing method by almost 70\% performance.

In future, we will apply the proposed schemes to a mobile edge-computing platform to provide multicriteria decision services to improve the performance of location-based recommendation applications.


%

\appendices




\ifCLASSOPTIONcaptionsoff
\newpage
\fi




\bibliographystyle{IEEEtran}
\bibliography{IEEEabrv,reference}

\begin{IEEEbiography}[{\includegraphics[width=1in,height=1.25in,clip,keepaspectratio]{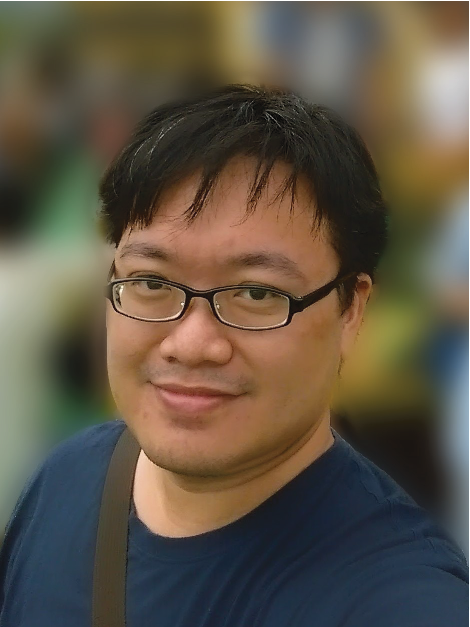}}]{Chuan-Chi Lai}
	(S'13 -- M'18) received his Ph.D. degree in computer science and information engineering from the National Taipei University of Technology, Taipei, Taiwan in 2017.
	
	Currently, Currently, he serves an Assistant Professor in the Department of Information Engineering and Computer Science, Feng Chia University, Taichung, Taiwan. His research interests include resource allocation, data management, information dissemination techniques, and distributed query processing over moving objects in emerging applications such as the Internet of Things, edge computing, aerial and mobile wireless applications.
	
	Dr. Lai has received the Postdoctoral Researcher Academic Research Award of Ministry of Science and Technology, Taiwan, in 2019, the Best Paper Awards in WOCC 2021 and WOCC 2018 conferences, and the Excellent Paper Award in ICUFN 2015 conference.
\end{IEEEbiography}

\begin{IEEEbiography}[{\includegraphics[width=1in,height=1.25in,clip,keepaspectratio]{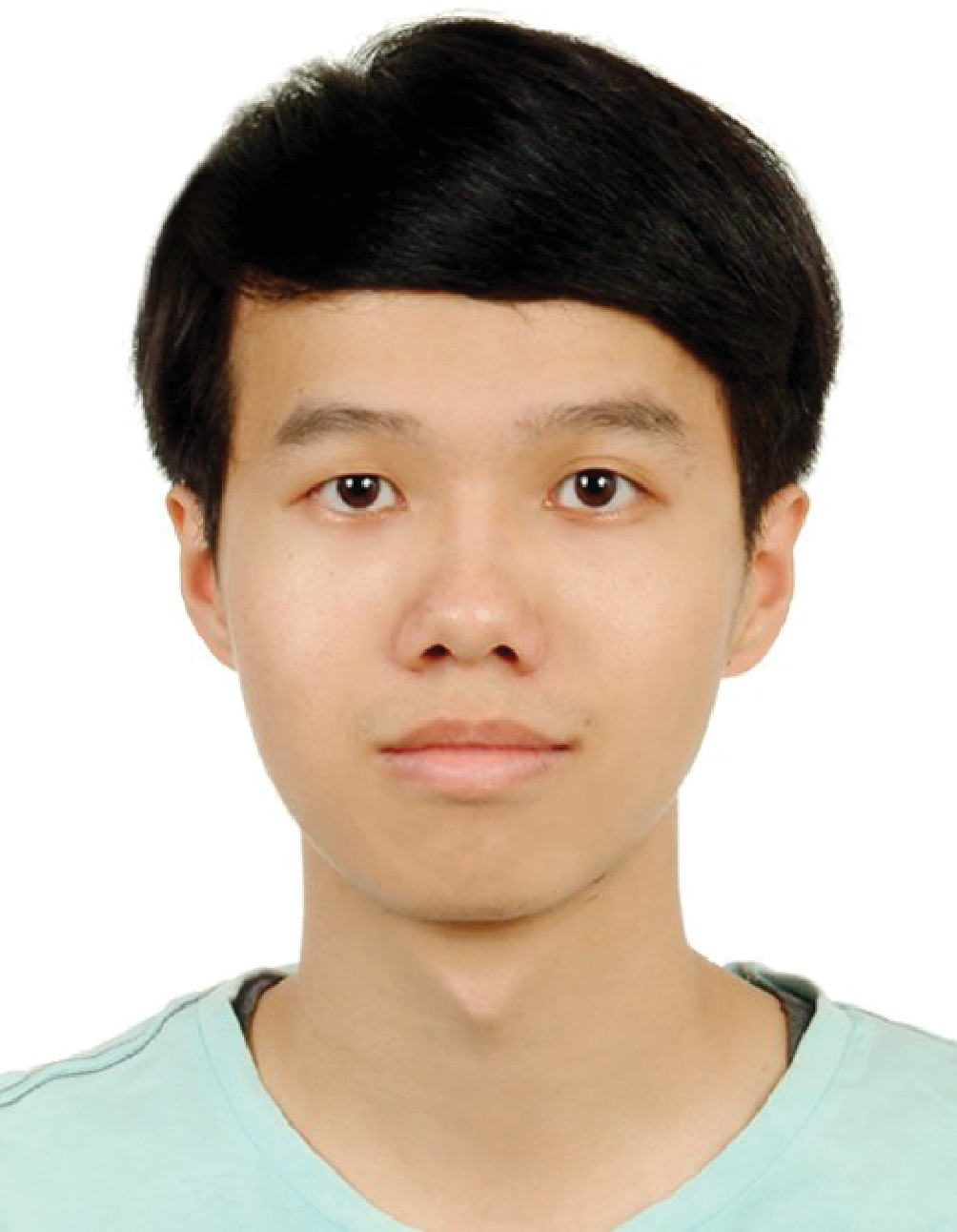}}]{Hsuan-Yu~Lin}
	received his MS degree from the Department of Computer Science and Information Engineering, National Taipei University of Technology (Taipei Tech), Taiwan in 2021. He joined the Applied Computing Laboratory in 2019 and was interested in developing data indexing and query processing algorithms for data analytic applications. Currently, he is an engineer with the Tool Productivity Optimization Department, Taiwan Semiconductor Manufacturing Co., Ltd. (TSMC), Hsinchu, Taiwan.
\end{IEEEbiography}

\begin{IEEEbiography}[{\includegraphics[width=1in,height=1.25in,clip,keepaspectratio]{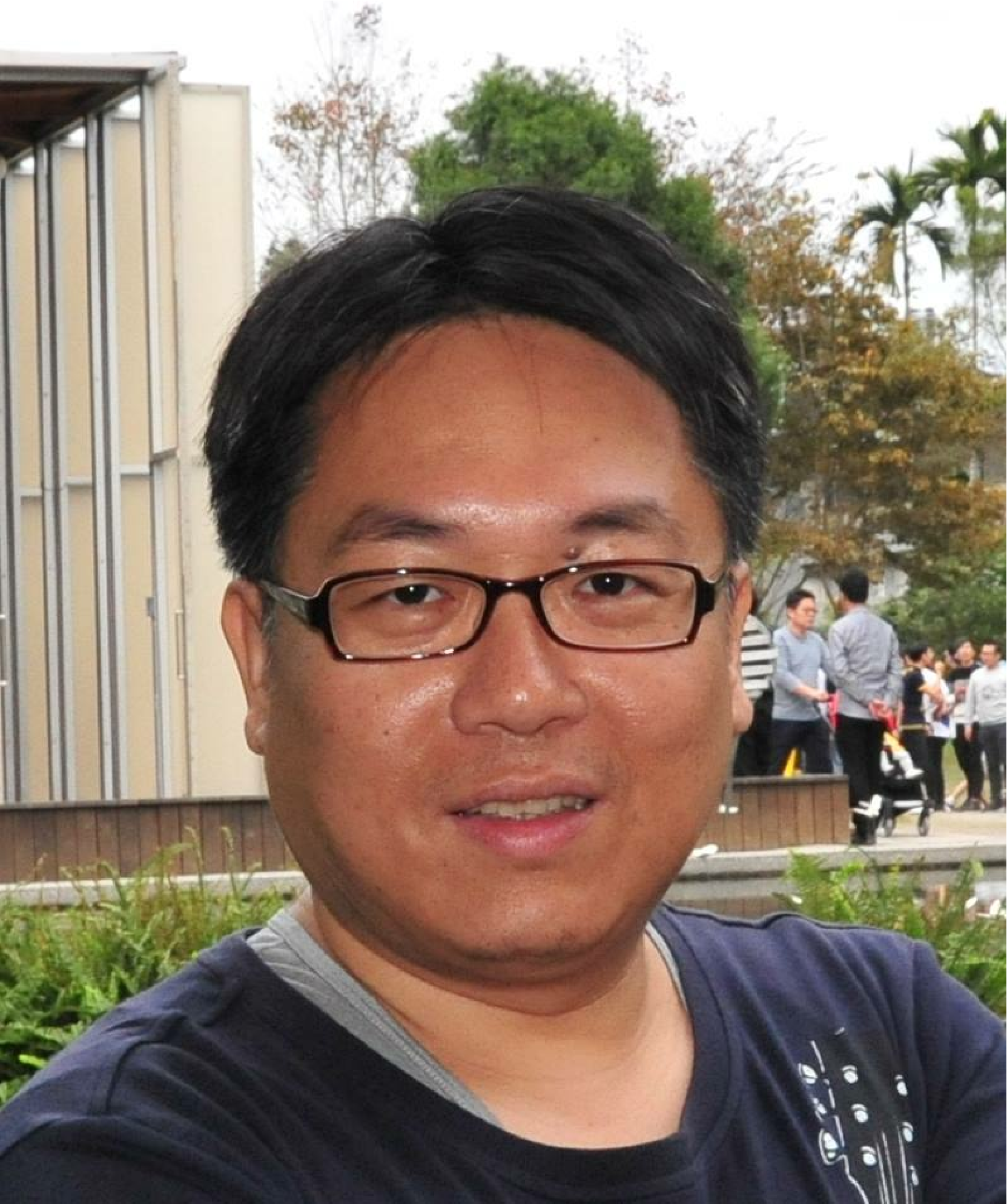}}]{Chuan-Ming Liu}
	(M'03) completed the Ph.D. in computer science from Purdue University, West Lafayette, IN, USA, in 2002. 
	Dr. Liu serves as a Professor in the Department of Computer Science and Information Engineering, National Taipei University of Technology (Taipei Tech), Taiwan. In 2010 and 2011, he held visiting appointments with Auburn University, Auburn, AL, USA and Beijing Institute of Technology, Beijing, China. In addition to his association with several journals, conferences, and societies, he has published more than 100 papers in numerous prestigious journals and international conferences. His research interests include big data management and processing, uncertain data management, data science, spatial data processing, data streams, ad hoc and sensor networks, and location-based services.
\end{IEEEbiography}

\end{document}